\documentclass[10pt]{article}

\usepackage{amsmath}
\usepackage{amssymb}

\usepackage{natbib}

\usepackage{graphicx}

\usepackage{color}
\usepackage{verbatim}

\usepackage[labelfont=bf,labelsep=period,justification=raggedright]{caption}

\makeatletter
\renewcommand{\@biblabel}[1]{\quad#1.}
\makeatother

\date{}

\begin{document}

% Title must be 150 characters or less
\begin{flushleft}
{\Large
\textbf{Error-prone polymerase activity causes multinucleotide mutations in humans}
}
% Insert Author names, affiliations and corresponding author email.
\\
Kelley Harris$^{1 \ast}$ 
and
Rasmus Nielsen$^{2,3,4}$
\\
\bf{1} Department of Mathematics, University of California Berkeley, Berkeley, CA, USA
\\
\bf{2} Department of Integrative Biology, University of California Berkeley, Berkeley, CA, USA
\\
\bf{3} Department of Statistics, University of California Berkeley, Berkeley, CA, USA
\\
\bf{4} Center for Bioinformatics, University of Copenhagen, Copenhagen, Denmark
\\
\end{flushleft}

\textbf{Corresponding author:}

Kelley Harris

1601 Allston Way, Berkeley, CA 94703 USA

916-205-2904

kharris@math.berkeley.edu

\vspace{0.5 cm}

\textbf{Running title:}

Error-prone polymerase activity causes MNM in humans

\vspace{0.5 cm}

\textbf{Keywords:}

multinucleotide mutation; transition/transversion ratio; low-fidelity polymerase; translesion synthesis; linkage disequilibrium

\newpage

% Please keep the abstract between 250 and 300 words
%\section*{Abstract}
\begin{abstract}
About 2\% of human genetic polymorphisms have been hypothesized to arise via multinucleotide mutations (MNMs), complex events that generate SNPs at multiple sites in a single generation. MNMs have the potential to accelerate the pace at which single genes evolve and to confound studies of demography and selection that assume all SNPs arise independently. In this paper, we examine clustered mutations that are segregating in a set of 1,092 human genomes, demonstrating that the signature of MNM becomes enriched as large numbers of individuals are sampled. We estimate the percentage of linked SNP pairs that were generated by simultaneous mutation as a function of the distance between affected sites and show that MNMs exhibit a high percentage of transversions relative to transitions, findings that are reproducible in data from multiple sequencing platforms and cannot be attributed to sequencing error. Among tandem mutations that occur simultaneously at adjacent sites, we find an especially skewed distribution of ancestral and derived dinucleotides, with $\textrm{GC}\to \textrm{AA}$, $\textrm{GA}\to \textrm{TT}$ and their reverse complements making up 27\% of the total. These mutations that have been previously shown to dominate the spectrum of the error-prone polymerase Pol $\zeta$, suggesting that low-fidelity DNA replication by Pol  $\zeta$ is at least partly responsible for the MNMs that are segregating in the human population.  We incorporate our findings into a mathematical model of the multinucleotide mutation process that can be used to correct phylogenetic and population genetic methods for the presence of MNMs.
\end{abstract}

\section*{Introduction}

One of the core challenges in evolutionary biology is to explain the distribution of mutations in time and space and harness this knowledge to make inferences about the past. When two DNA sequences have numerous differences that are spaced closely together, they are inferred to have been diverging for a relatively long time, the two lineages accumulating mutations at a steady rate since they diverged from their last common ancestor. In contrast, when two sequences have few differences that are spaced far apart, they are inferred to have diverged from a common ancestor relatively recently. This logic is the basis of a widely used class of methods that infer detailed demographic histories from the spacing between SNPs in a sample of whole-genome sequence data \citep{hobolth2007, li2011, sheehan2013, harris2013}.

To improve the accuracy of population genetic inference from the spacing between SNPs, it will be important to assess the validity of standard assumptions about the mutational process. One such assumption is that mutations occur independently conditional on the genealogical history of the data; however, there are numerous lines of evidence that 1--5\% of SNPs in diverse eukaryotic organisms are produced by multinucleotide mutation events (MNMs) that create two or more SNPs simultaneously. If simultaneously generated mutations are regarded as independent during population genetic analysis, the ages of the clustered variants will be overestimated. This could be important not only for the inference of demographic histories, but also for other endeavors such as the detection of long-term balancing selection. Closely spaced SNPs with ancient times to common ancestry can provide evidence that genetic diversity has been maintained by natural selection \citep{leffler2013, charlesworth2006, segurel2012}, and simultaneous mutations have the potential to distort or mimic these signals.

One line of evidence for MNM comes from \emph{de novo} mutations that occur in populations of laboratory organisms including \emph{Drosophila melanogaster} \citep{keightley2009, schrider2013}, \emph{Arabidopsis thaliana} \citep{ossowski2010}, \emph{Caenorhabditis elegans} \citep{denver2004, denver2009}, and \emph{Saccharomyces cerevisiae} \citep{lynch2008}, as well as \emph{de novo} mutations detected by looking at human parent-child-trios \citep{schrider2011}. Mutations occurring \emph{de novo} are often found clustered together in pairs within distances ranging from 2 bp to tens of kb; such clusters should be exceedingly rare under mutational independence. 

In yeast, there is additional evidence that MNMs are created by the activity of polymerase $\zeta$, an error-prone translesion polymerase that extends DNA synthesis past mismatches and damage-induced lesions \citep{sakamoto2007, stone2012}. Pol $\zeta$ is also responsible for MNMs that occur during somatic hypermutation of DNA that encodes the variable regions of mouse immunoglobulins \citep{daly2012, saribasak2012}. These results were established by inactivating Pol $\zeta$ in mutant yeast strains and adult mouse cells. However, it has not been possible to test experimentally whether Pol $\zeta$ creates heritable MNMs in higher eukaryotes. Pol $\zeta$ is required for rapid cell proliferation in normal embryonic development, making Pol $\zeta$ knockout mice embryonically inviable \citep{bemark2000, esposito2000, wittschieben2000}. 

Clusters of \emph{de novo} mutations are not the only line of evidence for heritable MNM in eukaryotes. Additional evidence for MNM  has been obtained by looking at patterns of linkage disequilibrium (LD) between older SNPs that segregate in natural populations. Schrider, \emph{et al}. and Terekhanova, \emph{et al}. examined pairs of nearby SNPs in phased human haplotype data and found that the two derived alleles occurred more frequently on the same haplotype than on different haplotypes \citep{schrider2011, terekhanova2013}. When two mutations occur independently, their derived alleles should occur on the same haplotype only 50\% of the time; in contrast, MNM should always produce mutation pairs with the two derived alleles on the same haplotype. Using a different counting argument, Hodgkinson and Eyre-Walker also concluded that many SNP pairs occurring at adjacent sites were generated by a simultaneous mutational mechanism \citep{hodgkinson2010}. They noted that adjacent linked SNPs outnumber SNPs 2 bp apart by a factor of two, when the two types of pairs should have equal frequency under the assumption of independent mutation.

To gather more data about the MNM process, it will be impractical to rely on \emph{de novo} mutations and essential to harness LD information. Although it is easiest to classify a pair of SNPs as an MNM when the mutations are observed \emph{de novo}, eukaryotes have low enough mutation rates that fewer than 1 MNM per genome is expected to occur each generation on average. Motivated by this, we use an LD-based approach to identify signatures of MNM in the 1000 Genomes Phase I data, a public repository of 1,098 phased human genomes \citep{1000genomes2012}. This repository is 100-fold larger than the datasets that were used for previously published studies of MNM, and its size confers a great deal of power to characterize the multinucleotide mutational spectrum. We use coalescent theory to show that a sample of 2,184 haplotypes should contain a much higher ratio of MNMs to single-mutation SNPs in high LD than smaller datasets do, providing increased statistical power to characterize MNMs, and this prediction is corroborated by our analysis of the 1000 Genomes data.

In accordance with earlier studies of MNM, we find that patterns of linkage disequilibrium (LD) between close-together SNPs are incompatible with mutational independence.  We show that it instead is well-described by a simple mixture of independent and MNMs. In addition, we are able to leverage the size of the 1000 Genomes dataset to make several novel discoveries about MNMs. First, MNMs are heavily enriched for transversions, with a transition: transversion ratio of about 1:1 in contrast to the 2:1 genomewide average. More importantly, we find that linked mutations in humans are enriched for the same allelic types recorded by Stone, \emph{et al.} in lines of yeast that have nucleotide excision repair (NER) deficiencies and thus rely heavily on Pol $\zeta$ for translesion synthesis \citep{stone2012}. These frequent MNMs include the dinucleotide mutations $\textrm{GA}\to\textrm{TT}$ and $\textrm{GC}\to\textrm{AA}$ as well as mutations at non-adjacent sites that produce homogeneous AA/TT derived allele pairs. Such patterns are unlikely to result from errors in the DNA sequencing process and instead suggest that normal human Pol $\zeta$ activity generates at least some of the same MNMs that are produced by Pol $\zeta$ in NER-deficient yeast \citep{stone2012}.

\section*{Results}

Simultaneous mutations can be observed directly when they occur \emph{de novo} in offspring that have been sequenced along with their parents. In addition, many more MNMs can be inferred from linkage disequilibrium (LD) in data from unrelated individuals. Schrider, \emph{et al}. previously invoked simultaneous mutations to explain LD patterns in a phased diploid genome, observing that SNPs less than 10 bp apart were disproportionately likely to have their derived alleles lie on the same haplotype \citep{schrider2011}. In the spirit of this approach, we looked at the prevalence of neighboring SNPs in the 1000 Genomes Phase I data that occur in perfect LD, meaning that the two derived alleles occur in the exact same subset of the 2,184 sequenced haplotypes. We hereafter define a pair of \emph{close LD SNPs} to be a pair occurring less than 100 bp apart in perfect LD. A few MNMs will be missed because of recombination between the mutated sites, but we estimate that fewer than 0.5\% of all MNMs spanning less than 100 bp will be disrupted in this way (see Section~S4of the supporting information). 

\textbf{Excess nearby SNPs in LD:}

We counted 35,620 pairs of close LD SNPs in the 1000 Genomes phase I data with both sites passing genotype quality control and with a consistent ancestral state identifiable from a human/chimp/orang/macaque reference alignment (see Methods). Simultaneous mutations should always create SNPs in perfect LD, but we also expect some independent mutations to create SNPs in perfect LD, and we quantified this expectation by simulating data under a Poisson process model of independent mutation and recombination implemented in \texttt{ms} \citep{hudson2002}. We simulated a total of $4.8\times 10^8$ bp from an alignment of 2,184 haplotypes under a realistic human demographic model \citep{harris2013} and recovered 36,991 close LD SNP pairs. For comparison, we also simulated $1.8\times 10^8$ bp of data under the standard neutral coalescent with constant effective population size $N=10,000$, recovering 36,202 close LD SNP pairs.

\begin{figure}[!ht]
\begin{center}
\includegraphics[width=4in]{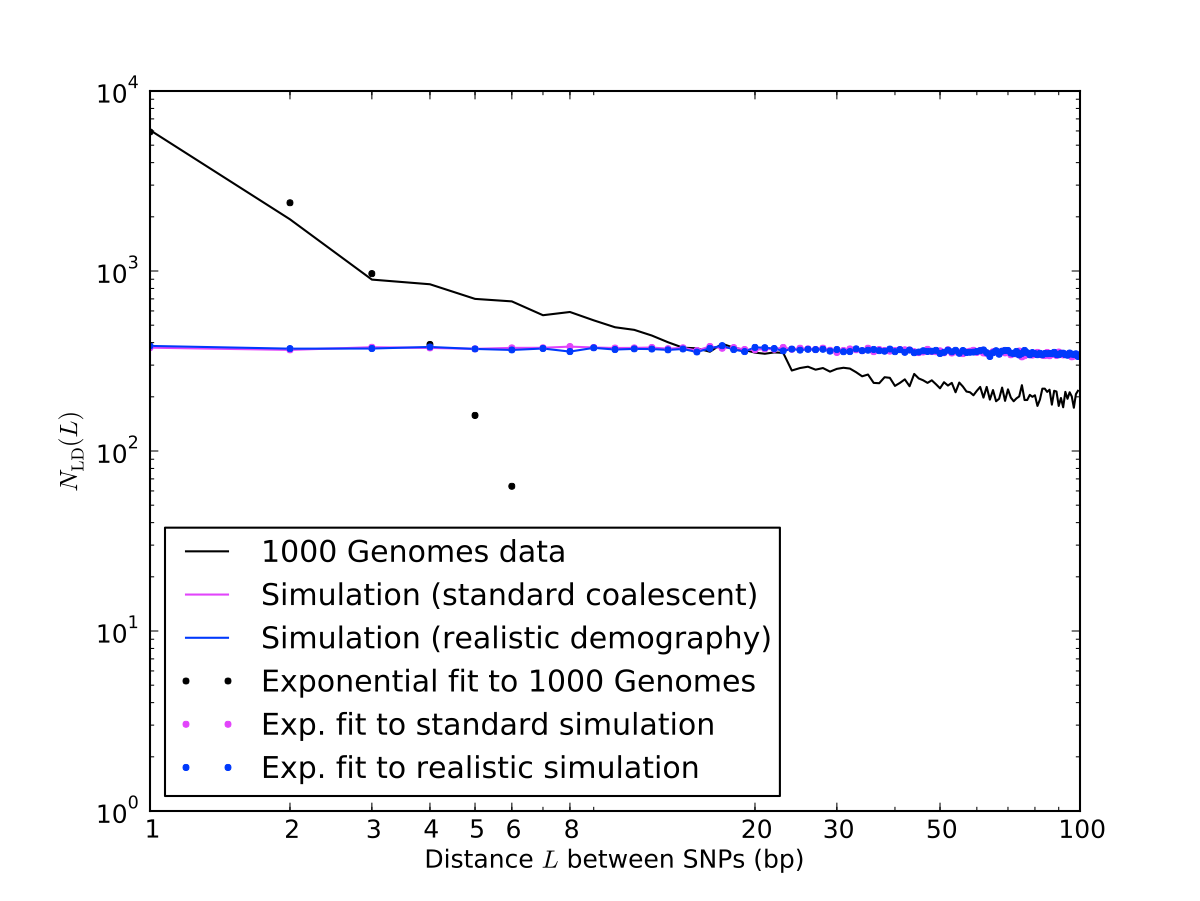}
\end{center}
\caption{
{\bf Nearby SNPs in LD: 1000 Genomes Phase I data vs. simulation under mutational independence.}  When we simulated 2,184 haplotypes under a realistic demographic model, we observed about 37,000 SNP pairs in LD separated by less than 100 bp in a sample of total length $4.8\times 10^8$ bp. Their spacing was distributed almost uniformly between 1 and 100 bp. Among these pairs, the spacing between SNPs was distributed almost uniformly. We observed much less uniformity in the distribution of distances between SNP pairs in LD in the 1000 Genomes data, with an extreme excess of SNPs in LD at 1--2 bp and a less extreme excess of SNPs at distances up to 20 bp apart. (Note that the axes are logarithmically scaled, making exponential curves appear concave downward).}
\label{LD_mixture_model}
\end{figure}

As shown in Figure~\ref{LD_mixture_model}, the distribution of distances between close LD SNPs were quite different in the simulated and in the real data, with the real data containing about 5-fold more adjacent SNPs in LD and a decaying excess of SNPs separated by up to 20 bp in LD. In contrast, there is comparatively little difference in the distances between SNPs in the data simulated under the realistic and the standard coalescent model, illustrating that the excess of SNPs in close LD cannot be attributed to assumptions regarding the demographic model.  Under the coalescent with independent mutation, the abundance of SNP pairs $L$ bp apart in LD should decline approximately exponentially with $L$ for small values of $L$ (see supporting information section S1), and we find this to hold for the simulated data in Figure~\ref{LD_mixture_model}.  

In contrast, the optimal least-squares exponential fit is a poor approximation to the abundance distribution of close LD SNP pairs in the 1000 Genomes data, which we denote by $N_{\textrm{LD}}(L)$. A possible explanation is that close linked SNPs are produced by a mixture of two processes, a point-mutation process that is accurately modeled by the coalescent and an MNM process that is not. 

\textbf{Closely linked SNPs have unusual transition/transversion frequencies:}

To our knowledge, no previous work has addressed whether MNMs have the same transition: transversion ratio as ordinary mutations. However, there is abundant evidence that different DNA polymerases produce mutations with different frequencies of ancestral and derived alleles. To investigate this question,  we measured the fractions of linked SNP pairs at distance $L$ that are composed of transitions, transversions, and mixed pairs (one transition plus one transversion). We denote these fractions $f_{\textrm{ts}}^{\textrm{LD}}(L),$ $f_{\textrm{tv}}^{\textrm{LD}}(L)$, and $f_{\textrm{m}}^{\textrm{LD}}(L)$. We also measured the analogous fractions $f_{\textrm{ts}}^{\textrm{non-LD}}(L),$ $f_{\textrm{tv}}^{\textrm{non-LD}}(L)$, and $f_{\textrm{m}}^{\textrm{non-LD}}(L)$ of transitions, transversions, and mixed pairs among SNPs not found in perfect LD. 

\begin{figure}[!ht]
\begin{center}
\includegraphics[width=4 in]{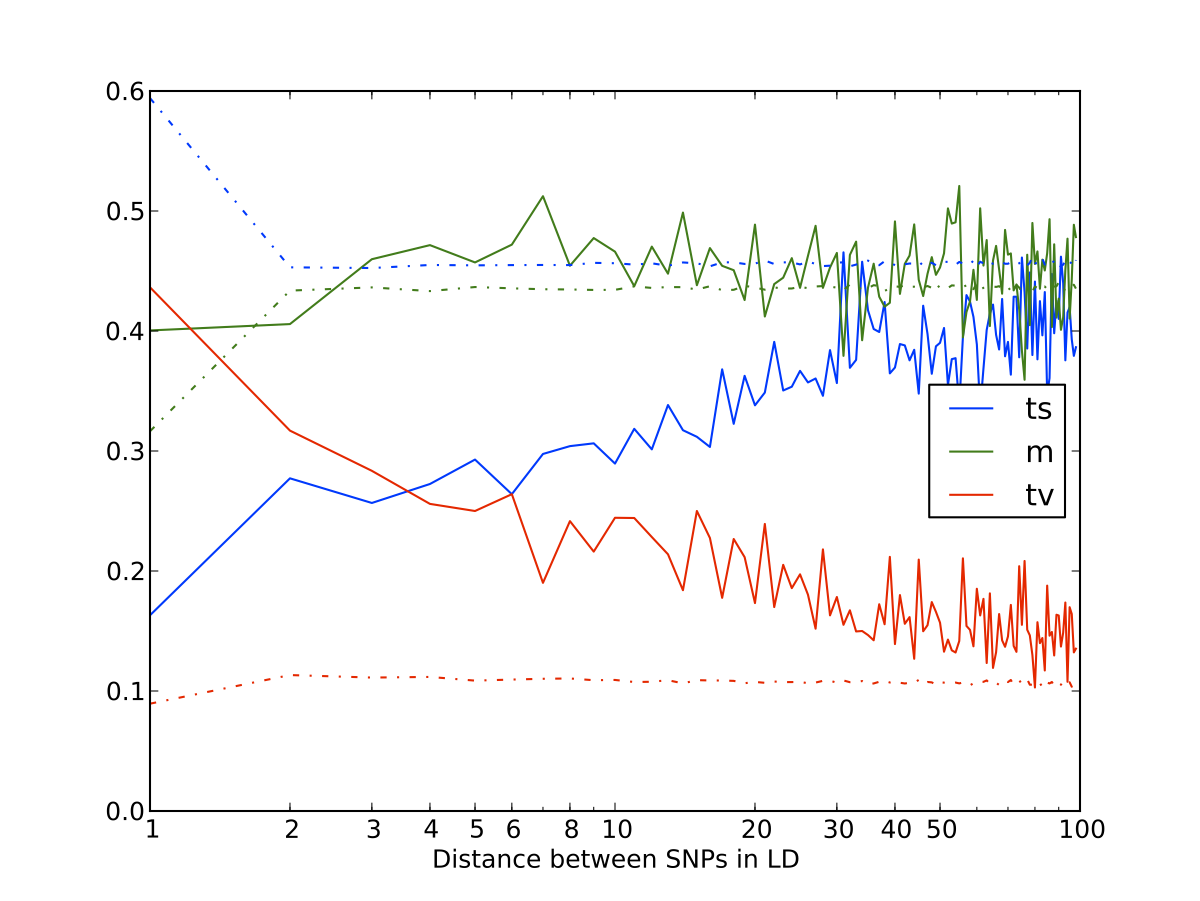}
\end{center}
\caption{
{\bf The relationship between LD and the transition:transversion ratio.} In this figure, the solid blue line plots the fraction of SNP pairs in LD that consist of 2 transitions. The fraction increases quickly as a function of the distance $L$ between SNPs, asymptotically approaching the fraction of SNP pairs not in LD that consist of 2 transitions (dashed blue line). The fraction of SNP pairs not in LD that consist of 2 transitions is nearly constant as a function of $L$ except for an excess of adjacent transition pairs resulting from double mutation at CpG sites. Although transversion pairs make up just over 10\% of unlinked SNP pairs, they account for over 40\% of adjacent SNPs in perfect LD and about 20\% of SNPs in LD at a distance of 10 bp apart.
}\label{transit_transv_freq_decay}
\end{figure}

In human genetic variation data, transitions are approximately twice as common as transversions \citep{kimura1980}. If the two mutation types of a SNP pair were chosen independently, we would therefore expect that $f_{\textrm{ts}}=0.66^2=0.44$, $f_{\textrm{tv}}=0.33^2=0.11$, and $f_{\textrm{m}}=2\times 0.66\times 0.33 = 0.45$. These predictions are very close to $f_{\textrm{ts}}^{\textrm{non-LD}}(L),$ $f_{\textrm{tv}}^{\textrm{non-LD}}(L)$, and $f_{\textrm{m}}^{\textrm{non-LD}}(L)$ for $L$ between 2 and 100. (Figure~\ref{transit_transv_freq_decay}). For $L=1$, $f_{\textrm{ts}}^{\textrm{non-LD}}(L)$ is larger than expected because of the elevated transition rate at both positions of CpG sites.

Among mutations in perfect LD, we found that $f_{\textrm{ts}}^{\textrm{LD}}(L)$, $f_{\textrm{tv}}^{\textrm{LD}}(L)$, and $f_{\textrm{m}}^{\textrm{LD}}(L)$ deviate dramatically from the expectation of mutational independence, adding support to the idea that many such SNPs are produced by a nonstandard mutational process. The frequency of transversion pairs declines with $L$; we found that 36.7\% of SNP pairs in LD at adjacent sites consisted of two transversions, compared to 11.1\% of SNP pairs in LD at a distance of 100 bp and 10.7\% of SNP pairs not in LD. These numbers are not just incompatible with a transition: transversion ratio of 2:1, but are also incompatible with two neighboring SNP types being assigned independently. If the SNP types were assigned independently, it should hold that $\sqrt{f_{\textrm{ts}}(L)}+\sqrt{f_{\textrm{tv}}(L)}=1$, an assumption that is violated for small values of $L$. We also found excess close LD transversions in human data sequenced by Complete Genomics (Supporting Figure~S1), suggesting that this pattern is not an artifact of the Illumina sequencing platform or the 1000 Genomes SNP-calling pipeline. 

\textbf{Estimating the fraction of perfect LD SNPs that are MNMs:}

Schrider, \emph{et al.} previously estimated the abundance of MNMs using the following analysis of a phased diploid genome: for distances $L$ ranging from 1 to 20 bp, they counted heterozygous sites $L$ bp apart where the derived alleles lay on the same haplotype and could potentially have arisen due to MNM. They compared this quantity, $S(L)$, to the number $D(L)$ of heterozygotes $L$ bp apart with the derived alleles on different haplotypes. If all mutations arise independently, $S(L)$ and $D(L)$ are expected to be equal, leading them to propose $S(L)-D(L)$ as an estimate of the number of MNMs spanning $L$ bp. We repeated this analysis on the 1000 Genomes data, subsampling each possible pair $H$ from among the 2,184 phased haplotypes. For each $L$ between 1 and 100 bp, we obtained counts $S^H_{\textrm{ts}}(L),$ $S^H_{\textrm{m}}(L),$ and $S^H_{\textrm{tv}}(L)$ of transitions, mixed pairs, and transversions $L$ bp apart where one haplotype carried the two ancestral alleles and the other haplotype carried the two derived alleles. Similarly, we obtained counts $D^H_{\textrm{ts}}(L),$ $D^H_{\textrm{m}}(L),$ and $D^H_{\textrm{tv}}(L)$ where the derived alleles occurred on opposite haplotypes of $H$. Adding up these counts over all haplotype pairs subsampled from the 1000 Genomes data, we obtained global counts $S_t(L)$ and $D_t(L)$ for each pair type $t$. The quantity $(S_{\textrm{tv}}(L)-D_{\textrm{tv}}(L))/(S(L)-D(L))$, a direct estimate of the fraction of MNMs that are transversion pairs, is consistently slightly higher than $f^{\textrm{tv}}_{\textrm{LD}}(L)$ (Supporting Figure S2), as expected if close linked SNP pairs are a mixture of MNMs and linked independent mutations. 

\begin{figure}[!ht]
\begin{center}
\includegraphics[width=4 in]{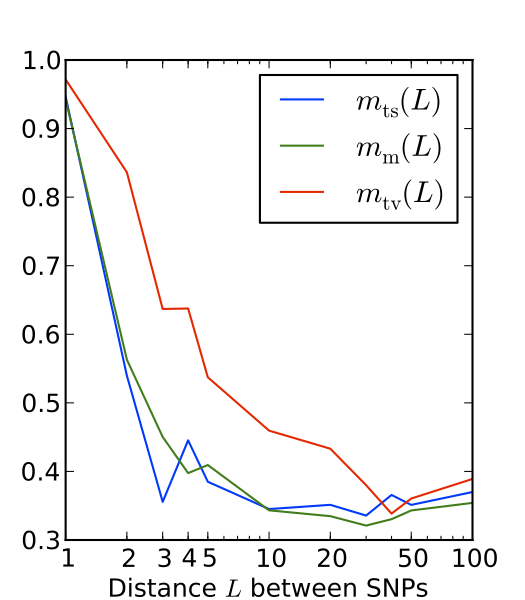}
\end{center}
\caption{
{\bf The fraction of SNPs in perfect LD caused by MNM.} The red curve plots our estimate of the fraction of SNPs in perfect LD at distance $L$ that were caused by simultaneous mutation. It is uniformly higher than our corresponding estimates for mixed pairs and transitions, plotted in green and blue.}
\label{percent_simul_muts}
\end{figure}

We were able to use $S_{\textrm{ts}}(L)-D_{\textrm{ts}}(L)$,  $S_{\textrm{m}}(L)-D_{\textrm{m}}(L)$, and  $S_{\textrm{tv}}(L)-D_{\textrm{tv}}(L)$ to  estimate the fraction of perfect LD SNPs that arose via MNM. Our simulations indicate that fewer than 1\% of MNMs 100 bp apart should be ultimately broken up by recombination (Supplementary Table S2); guided by this, we assume that MNMs are a subset of perfect LD SNP pairs. To make this assumption robust to phasing and genotyping error, we relax the definition of perfect LD to include site pairs where at most 2\% of samples carry a discordant genotype (see Methods). For each linked SNP pair, we count the number of subsampled haplotype pairs for which exactly one lineage contains the two derived alleles. Adding up these counts over all perfect LD SNPs, we obtain a count $S^{(\textrm{LD})}(L)$ that is strictly less than $S(L)$. We estimate that $m(L)=(S(L)-D(L))/S^{(\textrm{LD})}(L)$ is the fraction of perfect LD SNP pairs created by MNM. Similarly, $m_{\textrm{tv}}(L)=(S_{\textrm{tv}}(L)-D_{\textrm{tv}}(L))/S^{(\textrm{LD})}_{\textrm{tv}}(L)$ is the fraction of perfect LD transversions created by MNM. The results indicate that more than 90\% of SNPs in perfect LD at adjacent sites are MNMs (Figure \ref{percent_simul_muts}). At a distance of 5 bp between sites, 60\% of perfect LD transversions are predicted to be MNMs, in contrast to 40\% of perfect LD transitions and mixed pairs. At 100 bp between sites, about 35\% of perfect LD pairs appear to be MNMs, a figure that is similar across transitions and transversions. We calculate that MNMs spanning 1--100 bp account for 1.8\% percent of new point mutations (see Methods). Section S5 describes how to simulate data containing 1.8\% MNMs with realistic spacings of 1--100 bp. 

The 1000 Genomes data contains many SNP pairs that lie in perfect LD at distances of more than 100 bp apart. Although their transition/transversion ratios are close to the genomewide average, values of $m(L)$ suggest that more than 25\% of these are MNMs (Supplementary Table S1). Although MNMs spanning 10,000 bp appear to be rare events, 10-fold rarer than MNMs spanning only 100 bp, they appear only about 4-fold rarer than independent mutations occurring in perfect LD at 10,000 bp, making it possible to infer their distribution in the genome with high precision. 

\begin{figure}[!ht]
\begin{center}
\includegraphics[width=3 in]{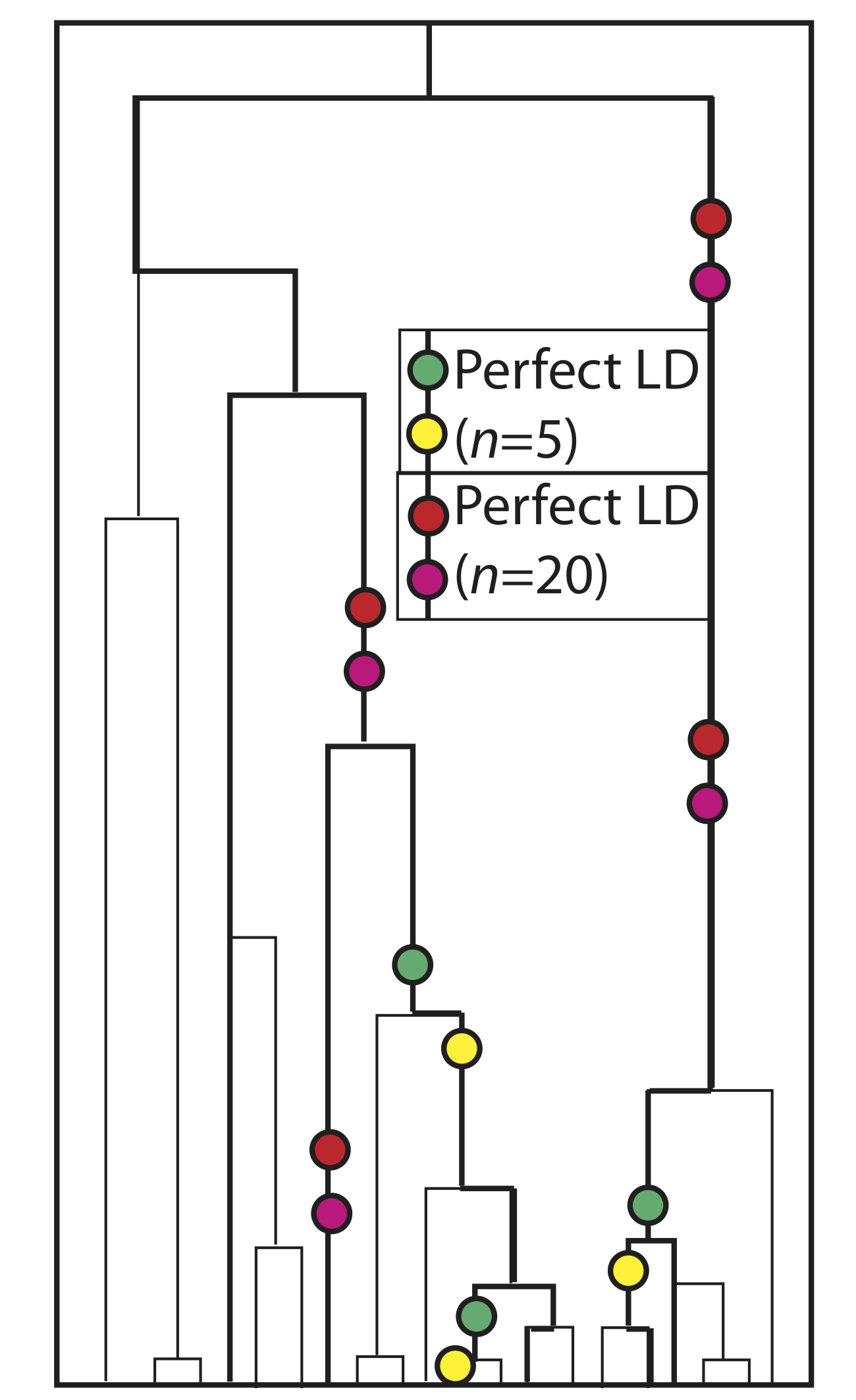}
\end{center}
\caption{
{\bf Independent mutations in perfect LD.} This figure depicts a 20-lineage coalescent tree with a 5-lineage subsample highlighted in bold. Green and yellow circles represent mutation pairs that appear in perfect LD only in the 5-lineage sample. In contrast, red and pink circles represent pairs of independent mutations that occur in perfect LD in the entire 20-lineage sample. These pairs are concentrated on the longest branches of the tree that are often ancestral to many lineages, making their site frequency spectrum enriched for high frequencies. }
\label{tree_branch_pic}
\end{figure}

The large sample size of the 1000 Genomes data not only ensures that a huge number of rare mutations can be observed, but also ensures that independent mutations occur in perfect LD much less often than in samples of fewer individuals. The reason for this is illustrated in Figure~\ref{tree_branch_pic}: if two mutations occurred at different time points on the genealogical tree of an entire population, sampling more individuals increases the probability of sampling one who carries the older mutation and not the younger one. To test this prediction, we counted SNP pairs that appear to be in perfect LD in smaller subsets of the 1000 Genomes data.  As proved in section S2 of the supporting information, the genealogies of large samples are dominated by shorter branches, on average, than the genealogies of smaller samples, implying that the percentage of perfected LD SNPs caused by MNM should be an increasing function of the number of sampled lineages. This implies that the abundance of transversions relative to transitions should also increase with the number of sampled lineages. 

\begin{figure}[!ht]
\begin{center}
\includegraphics[width=4 in]{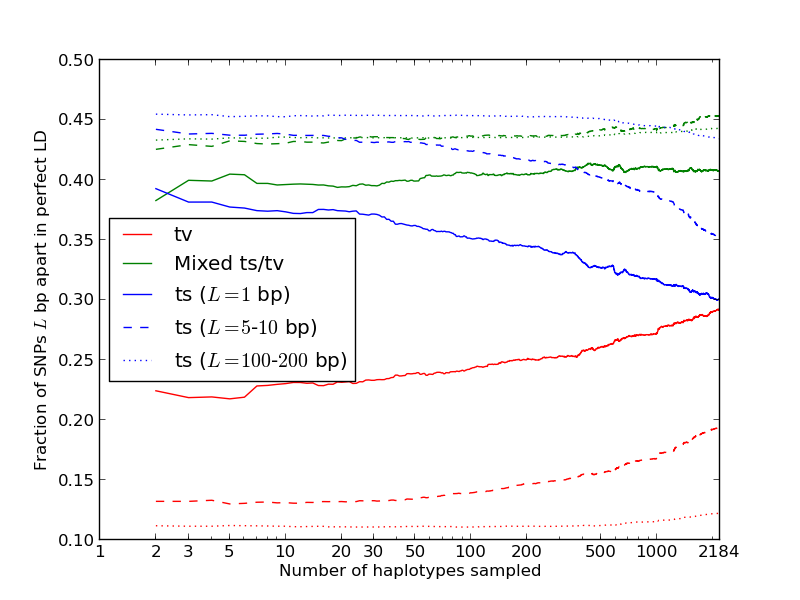}
\end{center}
\caption{
{\bf Enrichment of transversion pairs and MNMs with increasing sample size.} We generated subsamples of the 1000 Genomes data containing 2--2,184 haplotypes and computed the percentages of transversion pairs, transition pairs, and mixed pairs for perfect LD SNPs in each dataset. As the number of sampled haplotypes increases, the percentage of perfect LD SNPs that are MNMs should increase, leading to an increase in the frequency of transversions and a decrease in the frequency of transitions. This effect is most apparent when the SNPs are adjacent (1 bp apart) or very close (5--10 bp apart). However, perfect LD SNPs that lie 100--200 bp apart display the same pattern, indicating that MNMs spanning 100--200 bp are much less common but are still evident in samples of many lineages. }
\label{subsample_lineages}
\end{figure}

In pairs of adjacent perfect LD SNPs, we find that the percentage of transversion pairs increases very quickly with the number of lineages, making up 27\% of the total when only 2 haplotypes are sampled and nearly 40\% of the total when all 2,184 haplotypes are sampled (Figure~\ref{subsample_lineages}). This result is even more dramatic for transitions at CpG sites, where the rate of nonsimultanenous double mutations is elevated by deamination of methylated cytosine to thymine. When we count the fraction of adjacent SNPs in perfect LD that are of the type CG $\to$ TA as a function of the number $n$ of lineages sampled, it declines nearly 10-fold as $n$ increases from 2 to 2184 (see Supplementary Figure~S4). For perfect LD SNPs that occur 100--200 bp apart, the percentage of transversion pairs increases much more slowly than for adjacent perfect LD SNPs. However, it is still 10\% higher in a sample of 2,184 haplotypes than in samples of 2 to 1,000 haplotypes (Figure~\ref{subsample_lineages}). 

\textbf{Clustering of simultaneous mutations:}

Mutation-accumulation experiments have reported MNMs spanning long genomic distances \citep{denver2009, keightley2009, schrider2011, schrider2013}, and yeast studies have suggested a possible mechanism for their formation. Roberts, \emph{et al}. reported that double-strand breakage and subsequent repair can create sparse clusters of mutations spanning a megabase or more, with a mean spacing of 3,000 bp between simultaneous mutation events \citep{roberts2012}. We found evidence for higher-order mutational clustering by counting groups of mutations in perfect LD with fewer than 1000 bp between each adjacent pair and plotting the distribution of cluster size, which ranged from 2 to 46 SNPs. The distribution had a fatter tail than the distribution of perfect LD clusters in an equivalent amount of simulated data, where the largest cluster contained 23 perfect LD SNPs (Supplementary Figure S3). 

\textbf{The effect of complex mutation on the site frequency spectrum:}

\begin{figure}[!ht]
\begin{center}
\includegraphics[width=6 in]{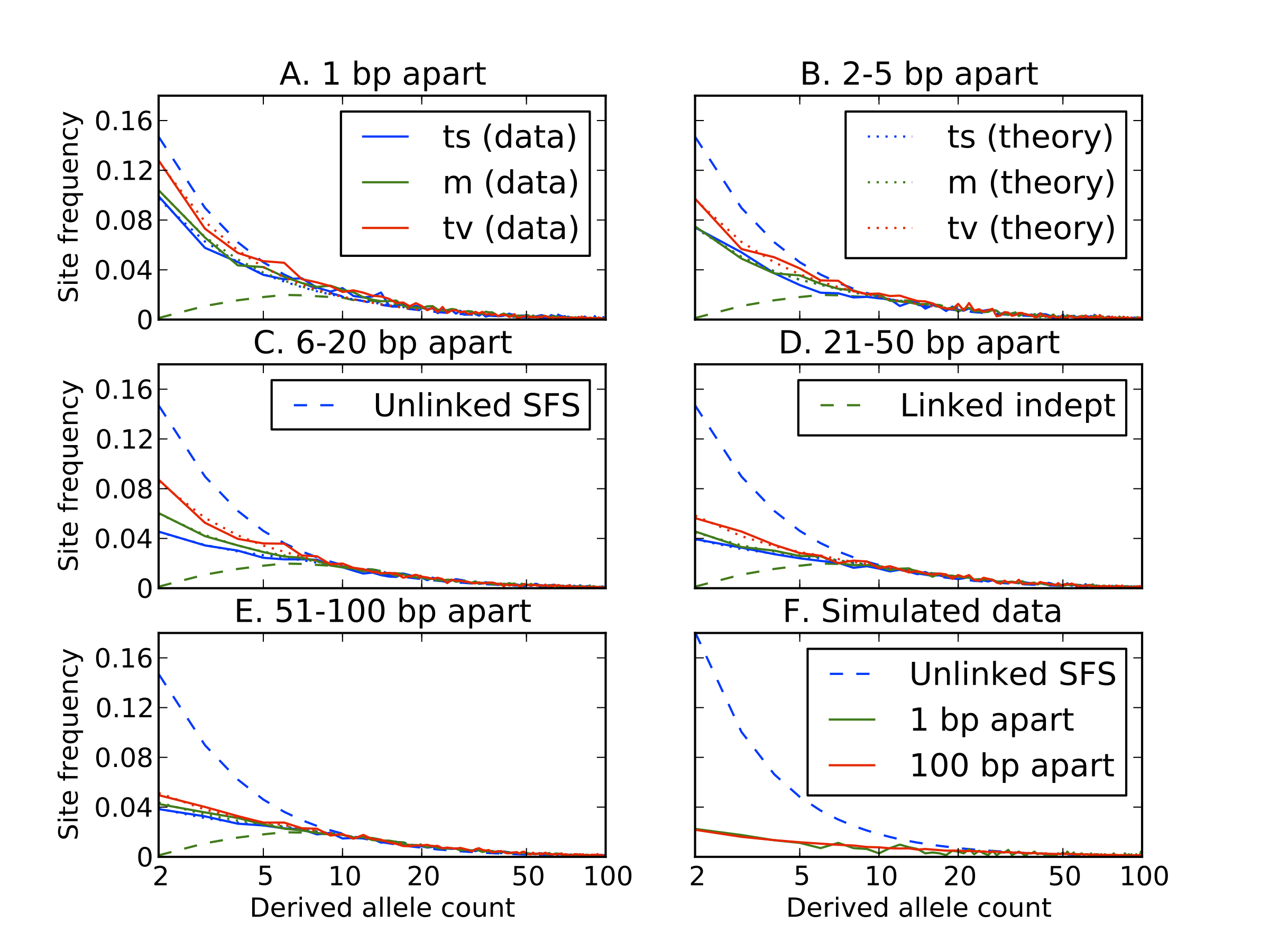}
\end{center}
\caption{{\bf Site frequency spectra of perfect LD mutations.} Each of the first five panels contains site frequency spectra of transitions, mixed pairs, and transversions found in perfect LD in the 1000 Genomes data. Singletons are excluded because they cannot be phased and therefore perfect LD status cannot be determined. For comparison, each panel contains the population-wide SFS of unlinked SNPs as well as the inferred SFS of linked independent mutations. SNP pairs are binned according to the distance between them, showing that close-together SNPs and transversions have spectra closer to the population SFS, while far-apart SNPs and transitions appear more weighted toward linked independent mutations. Dotted lines show the frequency spectra predicted by Equation~\eqref{SFS_inference} for each length and pair type category, assuming that the green dotted line depicts the correct SFS of linked independent mutations and that Figure~\ref{compare_MNM_estimates} shows the correct MNM percentages in each category. For comparison, panel F shows a population SFS and perfect LD frequency spectra obtained from data simulated under a human demographic model. In the simulated data, there is no difference between the frequency spectra of linked independent mutations that lie 1 bp apart versus 100 bp apart. 
} \label{SFS_vs_LD}
\end{figure}

In addition to showing that large samples contain fewer linked independent mutations than smaller samples, Figure~\ref{tree_branch_pic} illustrates that linked independent mutations should be enriched for high frequencies relative to the site frequency spectrum (SFS) of ordinary mutations. High frequency mutations tend to occur on the longest branches of a genealogical tree, whereas low frequency mutations are scattered across many short branches that are each less likely to be hit with two separate mutations. Simulations confirm that linked independent mutations are biased toward high frequencies, with 6-fold fewer singletons and doubletons than the SFS of the dataset they come from (Figure~\ref{SFS_vs_LD}F). In contrast, MNMs should have the same SFS as ordinary point mutations as long as they are not affected differently by natural selection.  

Given a mixture of simultaneous and independent mutations, the SFS should be a linear combination of the site frequency spectra of independent and simultaneous linked mutations. The more heavily the mixture is weighted toward independent mutations, the more the SFS should be skewed toward high frequencies. In agreement with our inference that MNMs contain a high percentage of transversions, we observe that perfect LD transversions have lower frequencies on average than other perfect LD SNP pairs. In addition, far-apart perfect LD SNPs have higher frequencies than close-together pairs on average (Figure~\ref{SFS_vs_LD}). 

\begin{figure}[!ht]
\begin{center}
\includegraphics[width=4 in]{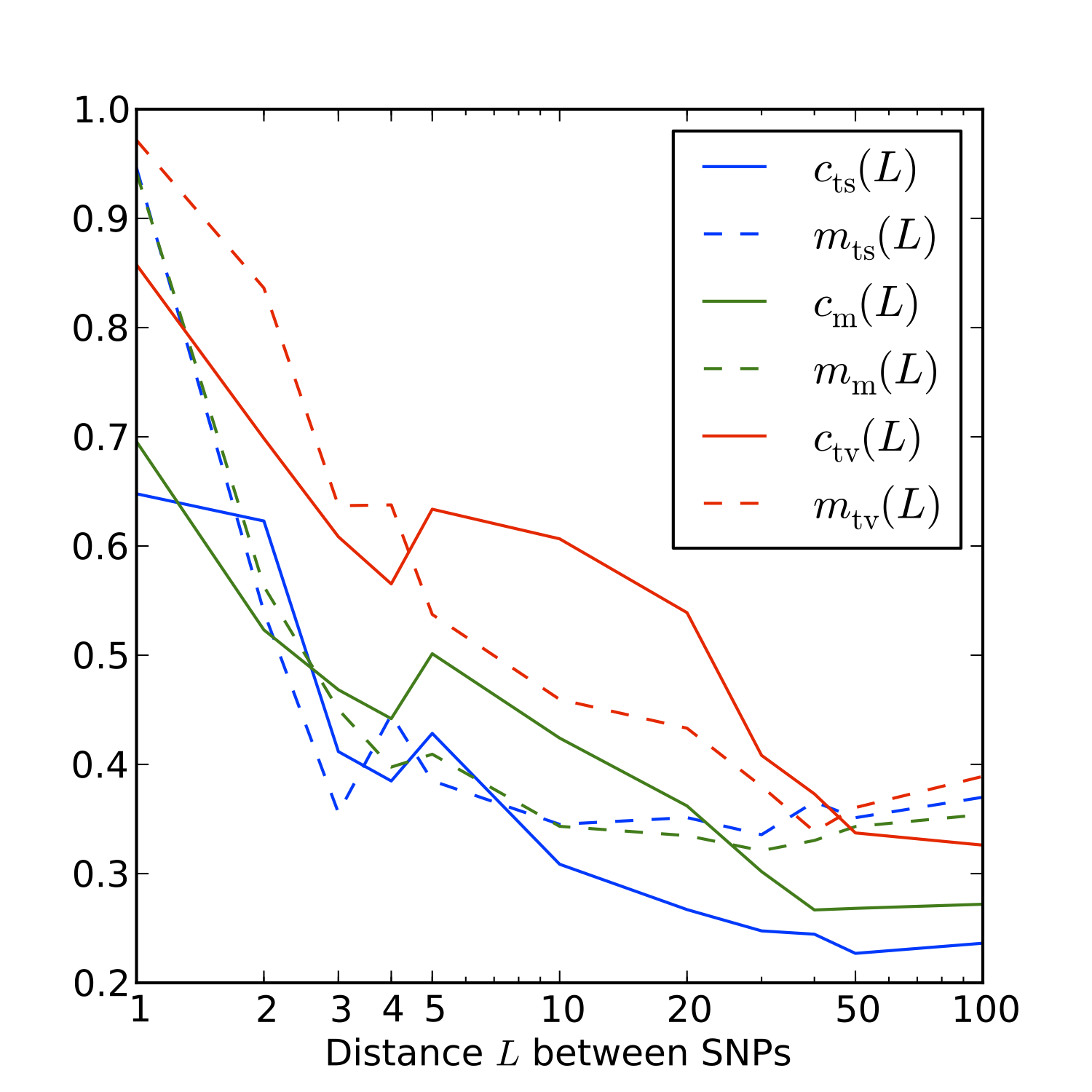}
\end{center}
\caption{{\bf Two estimates of MNM prevalence.} Here, the solid lines plot $c_t(L)$, our SFS-based estimate of the fraction of perfect LD mutations caused by MNM. For comparison, dotted lines plot the estimate $m_t(L)$ that is based on the excess of same-lineage derived alleles over different-lineage derived alleles in subsampled haplotype pairs.
} \label{compare_MNM_estimates}
\end{figure}

Using the empirical spectra of linked versus unlinked mutations, we devised a second method for estimating the fraction of perfect LD SNPs that are MNMs. For each mutation pair type (ts/m/tv), we compute the site frequency spectrum $\mathbf{S}(L)$ of perfect LD SNPs $L$ bp apart. We also computed a SFS $\mathbf{S}_{\textrm{global}}$ from the entire set of SNPs in the sample. It is not possible to measure the spectrum $\mathbf{S}_{\textrm{indept-LD}}$ of linked independent mutations directly, and so we numerically optimized the entries of this spectrum jointly with mixture model coefficients $c_{\textrm{ts}}(L), c_{\textrm{m}}(L), $ and $c_{\textrm{tv}}(L)$ between 0 and 1, one for each distance $L$ and mutation pair type $t$. We treated all entries of $\mathbf{S}_{\textrm{indept-LD}}$ as unknown free parameters and used the BFGS algorithm to minimize the following squared error residual $\mathbf{D}$:
\begin{equation}\label{SFS_inference}
\mathbf{D} = \sum_{i=2}^n\sum_{t\in \{\textrm{ts, m, tv}\}} \left( c_t(L)\times \mathbf{S}_{\textrm{global}}[i]+(1-c_t(L))\times \mathbf{S}_{\textrm{indept-LD}}[i]-\mathbf{S}_t(L)[i]\right)^2
\end{equation}
This has the effect of fitting each spectrum $\mathbf{S}_t(L)$ to the linear combination $c_t(L)\times \mathbf{S}_{\textrm{global}}+(1-c_t(L))\times \mathbf{S}_{\textrm{indept-LD}}$ of MNMs and linked independent mutations. The sum over $i$ starts at 2 to exclude singletons because they cannot be phased. Assuming that $\mathbf{S}_{\textrm{global}}$ is a good estimate of the SFS of MNMs, we take $c_t(L)$ to be an estimate of the fraction of MNMs among linked SNPs of type $t$ at distance $L$. In this way, we obtain estimates similar to the $m_t(L)$ estimates that we obtained earlier by measuring the excess of same-lineage of derived alleles (Figure~\ref{compare_MNM_estimates}). We find that $c_t(L)$ is larger than $m_t(L)$ for $L<3$ and $L>50$, but smaller than $m_t(L)$ at intermediate distances. The discrepancy might stem from simple noise in the data, but might also mean that processes like context-dependent mutation and strand bias create excess same-lineage pairs that do not lie in perfect LD.

\textbf{Evidence for error-prone synthesis by Polymerase $\zeta$:}

One mechanism that is known to generate MNMs \emph{in vivo} is error-prone lesion bypass by Polymerase $\zeta$, an enzyme found in all eukaryotes with the unique ability to extend primers with terminal mismatches \citep{gan2008, waters2009}. At a replication fork that has been stalled by a lesion, Pol $\zeta$ is responsible for adding bases to the strand containing the lesion and then extending replication for a few base pairs before detaching and allowing a high-fidelity enzyme to resume synthesis. During this extension phase, it has the potential to create clustered errors. Experimental work in yeast has confirmed that Pol $\zeta$ generates MNMs \citep{sakamoto2007, stone2012}, and the same enzyme has been linked to somatic hypermutation in the MHC \citep{daly2012, saribasak2012}. 

Translesion synthesis by Pol $\zeta$ is not the only pathway that has the potential to create MNMs. Eukaryotes utilize at least seven different DNA replication enzymes that are considered ``error-prone" \citep{goodman2002, waters2009} and have mutation spectra with low transition/transversion ratios \citep{mcdonald2011}. However, we specifically analyzed human MNMs for signatures of Pol $\zeta$ activity because a unique dataset was available to make this effort possible. Specifically, we were able to compare linked adjacent mutations in the 1000 Genomes data to tandem (adjacent) mutations recorded from a yeast strain bred by Stone, \emph{et al}. to be deficient in nucleotide excision-repair machinery and rely heavily on Pol $\zeta$ to bypass lesions that stall replication forks  \citep{stone2012}. Stone, \emph{et al}. recorded a total of 61 spontaneous tandem mutations; these were even more heavily weighted toward transversions than linked SNPs in the 1000 Genomes data, with 52.5\%  transversion pairs, 37.7\% mixed pairs, and only 9.8\% transition pairs. 

\begin{figure}
\begin{center}
\includegraphics[width=6 in]{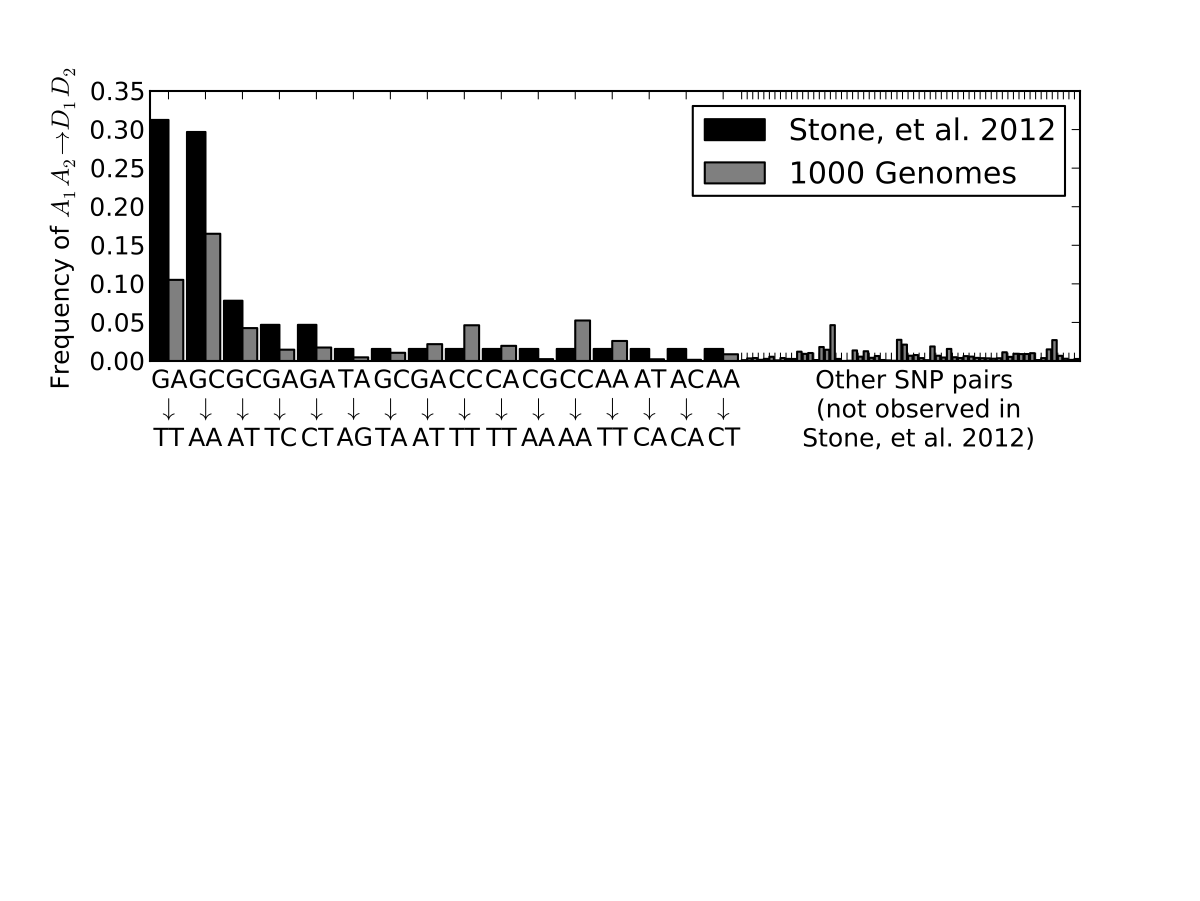}
\end{center}
\caption{
{\bf Tandem mutations caused by Pol $\zeta$.}  Blue bars plot the frequencies of specific tandem mutations observed by Stone, \emph{et al}. in yeast deficient in nucleotide-excision repair machinery. Each mutation type is pooled with its reverse complement because there is no way to know on which DNA strand a mutation occurred. The two mutations $\textrm{GC}\to\textrm{AA}$ and $\textrm{GA}\to\textrm{TT}$ account for more than 60\% of all tandem mutations observed by Stone, \emph{et al}. As shown in red, these are also the two most common types of mutations occurring at adjacent sites of the 1000 Genomes data in perfect LD.
}\label{stone_mutations}
\end{figure}

Two particular tandem mutations composed more than 60\% of the tandem mutations in the Stone, \emph{et al}. yeast. One of them, $\textrm{GA}\to \textrm{TT},$ is a transversion pair that made up 31\% of the total. The other, $\textrm{GC}\to\textrm{AA}$, is a mixed pair that made up 30\% of the total. We found that these were also by far the most common adjacent linked SNPs in the 1000 Genomes data, with $\textrm{GC}\to\textrm{AA}$ comprising 16\% of the total and $\textrm{GA}\to\textrm{TT}$ comprising 11\%. No other single mutation type accounts more than 5\% of the linked adjacent mutations in the 1000 Genomes data, and no other type accounts for more than 7\% of the Stone, \emph{et al}. tandem mutations (Figure~\ref{stone_mutations}). 

In addition to 61 tandem mutations affecting adjacent base pairs, Stone, \emph{et al}. recorded 210 complex mutations where two or more substitutions, insertions, and/or deletions occurred at non-adjacent sites within a single 20 bp window. From this dataset, we extracted 84 pairs of simultaneous substitutions at distances of 2--14 bp apart. These pairs had almost the same transition/transversion makeup as the tandem substitutions, being comprised of 53.6\% transversions, 36.9\% mixed pairs, and 9.5\% transitions. 

\begin{figure}
\begin{center}
\includegraphics[width=4 in]{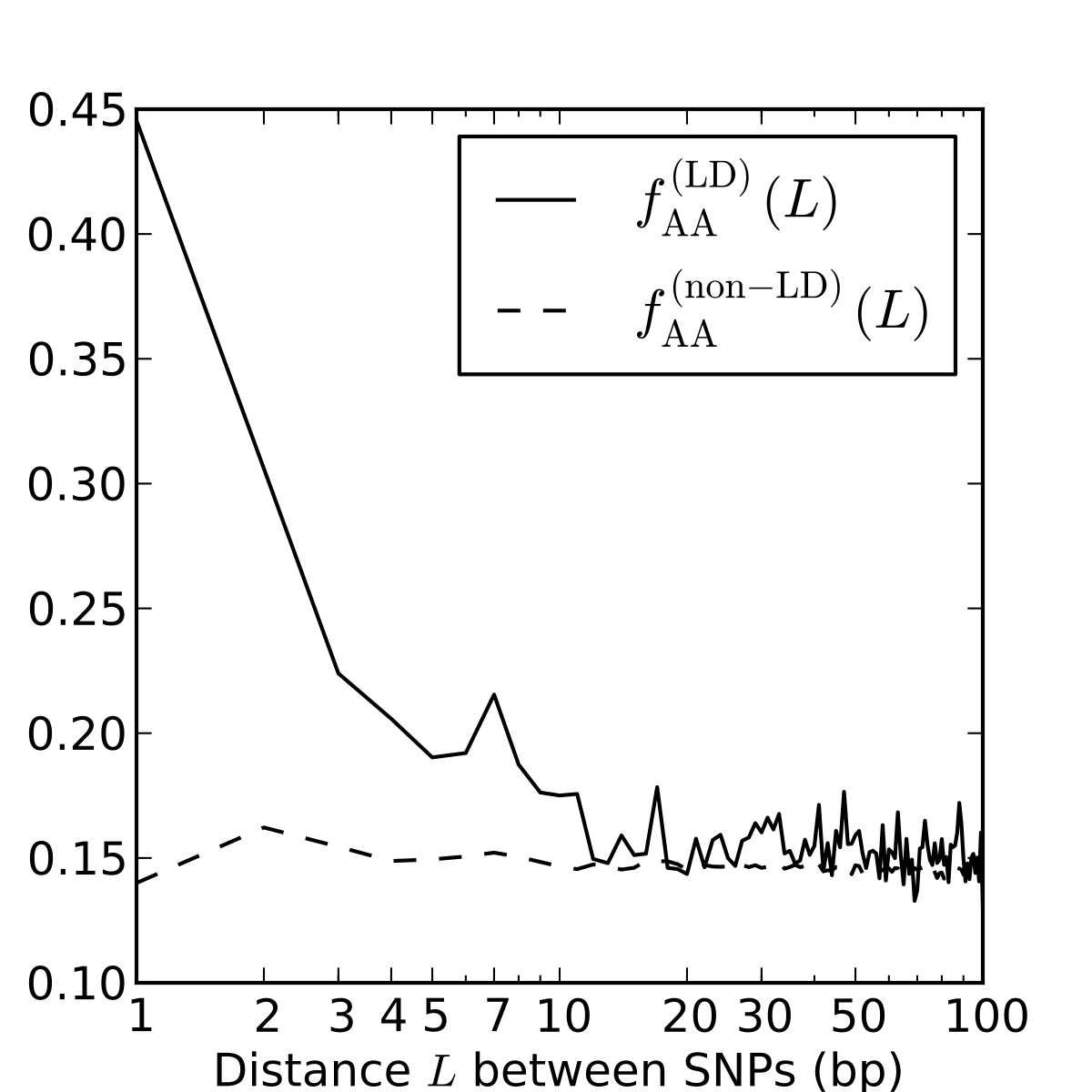}
\end{center}
\caption{
{\bf Linked derived AA/TT allele pairs in the 1000 Genomes data.}  After observing that a high fraction of yeast MNMs had homogeneous AA/TT derived allele pairs, we tabulated the frequencies $f_{\textrm{AA}}^{\textrm{(LD)}}(L)$ of AA/TT derived allele pairs among perfect LD SNPs $L$ bp apart in the 1000 Genomes data. Here, $f_{\textrm{AA}}^{\textrm{(LD)}}(L)$ is plotted alongside the theoretical estimate $\hat{f}_\textrm{AA}^{\textrm{(LD)}}(L)$ from equation~\eqref{AA_freq_fit}, which posits that MNMs produce AA/TT derived allele pairs 2--4 times as often as independent mutation pairs do. For comparison, we also plot $f_{\textrm{AA}}^{\textrm{(non-LD)}}(L)$, the frequency of AA/TT derived allele pairs among SNPs not in perfect LD. This fraction is consistently lower than $f_{\textrm{AA}}^{\textrm{(LD)}}(L)$ and does not decrease with the distance between SNPs. 
}\label{frac_AA}
\end{figure}

Among the non-adjacent yeast mutation pairs, $\textrm{GA}\to \textrm{TT}$ and $\textrm{GC}\to\textrm{AA}$ were not particularly common, making up only 4.8\% and 1.2\% of the total, respectively. However, 44.0\% of the derived allele pairs were ``AA" or ``TT" (compared to 72.1\% of adjacent mutation pairs). This percentage is much higher than what we would expect in two mutations that occurred independently. Mutation accumulation studies have shown that 33\% of yeast mutations have derived allele A (by A/T symmetry, 33\% also have derived allele T) \citep{lynch2008}. From this, we expect the fraction of AA/TT derived allele pairs to be only $2\times 0.33^2=0.22$. We found that AA and TT were similarly overrepresented among the derived allele pairs in linked human SNPs. In Figure~\ref{frac_AA}, we plot the fraction $f_{\textrm{AA}}(L)$ of derived AA/TT allele pairs as a function of the distance $L$ between perfect LD SNPs, charting its decline from $f_{\textrm{AA}}(1)=0.445$ through $f_{\textrm{AA}}(100)=0.144$. 

In their 2011 study of MNMs in human trios, Schrider, \emph{et al.} tabulated frequencies of all possible 144 dinucleotide substitutions but did not report excess AA/TT derived allele pairs or Pol $\zeta$-associated mutations of the types $\textrm{GA}\to \textrm{TT}$ or $\textrm{GC}\to\textrm{AA}$. We believe that these results differ because of our theoretical result that sampling more lineages enriches the ratio of true MNMs to linked independent mutations. To verify this, we replicated the Schrider, \emph{et al.} mutation frequency analysis on the 1000 Genomes data and obtained results that were similar to theirs (section S3 of the supporting information). We also found that the excess of Pol $\zeta$-associated mutations increases as more lineages are sampled (Supporting Figure S5), just as all perfect LD transversions increase in frequency with sample size (Figure~\ref{subsample_lineages}). Other AA/TT derived allele pairs increase in frequency as more lineages are sampled when considering SNPs less than 4 bp apart. Since this effect is not discernible for $L> 4$, derived AA/TT pairs might only play a significant role in closely spaced MNMs.

\textbf{Correcting downstream analyses for multinucleotide mutation:} 

As evidenced by Figures~\ref{LD_mixture_model} and~\ref{SFS_vs_LD}, MNMs can have considerable impact on summary statistics like the site frequency spectrum and the prevalence of linkage disequilibrium. These summary statistics provide clues about the genealogical histories of datasets and can be leveraged to infer demographic history, natural selection, population structure, recombination rates, and other quantities of interest. However, accurate inference depends on accurately modeling the process that generates data, and most population genetic models omit MNMs. 

\begin{figure}[!ht]
\begin{center}
\includegraphics[width=4in]{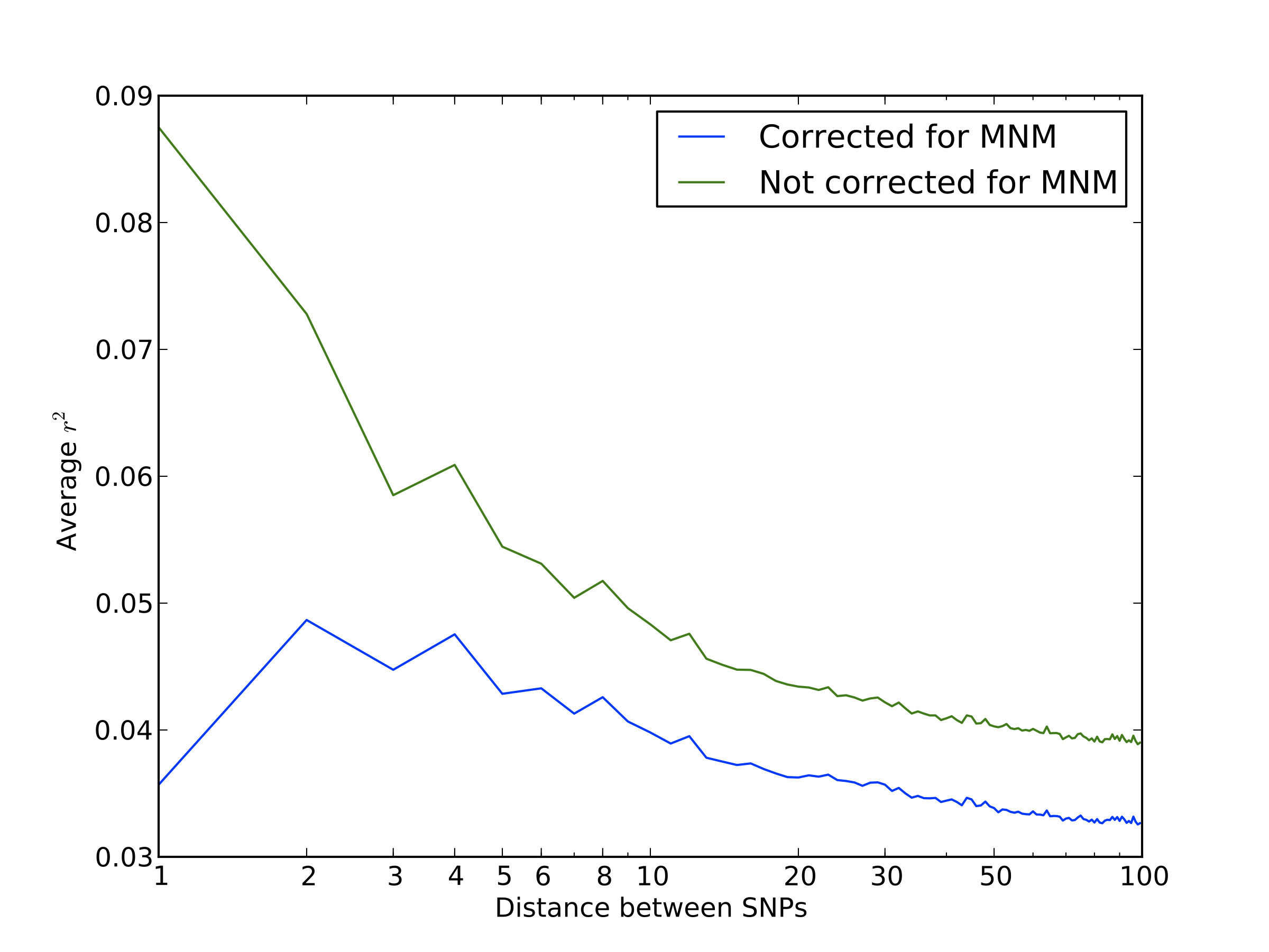}
\end{center}
\caption{{\bf Average $r^2$ LD correlations between 1000 Genomes SNP pairs.} The correlation coefficient $r^2$ between allele frequencies at neighboring sites is often used to measure the decay rate of genealogical correlation with genomic distance. However, we have seen that multinucleotide mutation creates excess LD compared to the expectation under independent mutation. We computed the average $r^2$ across all SNP pairs $L$ bp apart on chromosome 22, then corrected this value for the presence of MNM. $r^2_{\textrm{MNM}}$ is lower at a distance of 1 bp than a distance of 2 bp because of double deaminations at CpG sites that occur on separate lineages.
}\label{corrected_rsquared}
\end{figure}

One strategy for improving the accuracy of downstream analyses without adding much to their complexity is to identify  MNMs in a probabilistic way and remove them from the data. For each pair of SNPs occurring in perfect LD, we can estimate the probability that they were caused by an MNM as a function of their inter-SNP distance and transition/transversion status, then use this information to correct summary statistics for the presence of MNMs. To illustrate, we devise a method for correcting the correlation coefficient $r^2(L)$ that is commonly used to measure linkage disequilibrium as a function of genomic distance $L$ \citep{hill1968}. We computed $r^2(L)$ in the 1000 Genomes data as described in the methods and then devised a corrected statistic $r^2_{\textrm{MNM}}(L)$ that accounts for MNMs and estimates the average correlation between independent mutations. As shown in Figure~\ref{corrected_rsquared}, $r_{\textrm{MNM}}^2(L)$ is significantly less than $r^2(L)$ at short genomic distances.

\section*{Discussion}

We have uncovered a strong signature of multinucleotide mutation in 1,092 genomes sequenced by the 1000 Genomes consortium, with a large excess of close LD SNPs that cannot be explained by demography or mutational hotspots. This is consistent with earlier reports of MNM in smaller human datasets; however, MNMs are enriched relative to independent linked SNPs as more lineages are sampled and mutations are localized to increasingly short genealogical branches. 

By looking at the allelic composition of close LD SNPs containing MNMs, we found several signatures that are consistent with error-prone lesion bypass by Polymerase $\zeta$. One signature is an excess of transversions, the second is an excess of the dinucleotide mutations $\mathrm{GA}\to\mathrm{TT}$ and $\mathrm{GC}\to\mathrm{AA}$, and the third is a bias toward homogeneous AA/TT derived allele pairs. It remains an open question what percentage of human MNMs are introduced by Pol $\zeta$ and how many other DNA damage and repair mechanisms come into play. However, it is interesting that Pol $\zeta$ appears to create the same mutation types in the human lineage that it creates in yeast with artificial excision repair deficiencies. We are hopeful that MNM can be understood more completely in the future by comparing perfect LD SNPs to \emph{de novo} mutations from other sources. 

An important alternative hypothesis for the observed patterns is DNA sequencing or assembly errors in the 1000 Genomes data, but there are several different lines of evidence that show that our results cannot by explained by such errors.  First, we observed similar patterns in data sequenced by Complete Genomics using non-Illumina technology.  Secondly, the excess close LD SNPs that are enriched for transversions and AA/TT derived alleles are not only singleton mutations, but occur at a range of higher allele frequencies.  Errors could only cause such patterns if they occurred in an identical fashion in multiple individuals, mimicking the frequency distribution expected for mutations.  Thirdly, as already noted, the MNMs we infer are enriched for the same types as MNMs that were observed \emph{de novo} in yeast. The patterns we observe are consistent with MNM patterns that have been previously found using Sanger sequencing and other high-fidelity variant detection methods \citep{drake2005, levy2007, lynch2008, chen2009}.

 Most commonly used methods for analyzing DNA  sequences assume that mutations occur independently of each other.  The fact that this assumption is violated in human data, and perhaps most other eukaryotic data, may have a strong effect on many methods of inference.  Methods based solely on counting mutations, such as SFS based methods \citep{gutenkunst2009} will probably be minimally affected and mostly in their measures of statistical confidence.  In contrast, methods that explicitly use the spatial distributions of mutations, such as methods based on the distribution of the number of mutations in short fragments of DNA \citep{yang1997, wang2010, nielsen2001, gronau2011} should be strongly affected. Several recently developed methods analyze genomic data by explicitly modeling the spatial distribution of independent mutations  \citep{hobolth2007, li2011, sheehan2013, harris2013}, and these are at risk for bias in regions where SNPs are close together. However, confounding of these methods by MNM can be minimized by analyzing only a few individuals at a time and by disregarding pairs of SNPs less than 100 bp apart, which is often coincidentally done for the sake of computational efficiency \citep{li2011, harris2013}. MNMs likely have a stronger effect on methods that look at data from many individuals across short, allegedly non-recombining genomic fragments that are only 1 kb long and contain many SNPs fewer than 100 bp apart \citep{yang1997, gronau2011}.  However, our results can be used to devise bias-correction strategies, because as illustrated in Figure~\ref{percent_simul_muts}, it is straightforward to estimate the probability that a given pair of linked SNPs is an MNM. This also has the potential to improve the accuracy of phylogenetic tree branch length estimation and molecular-clock-based inferences, as well as dN/dS estimation, and their associated measures of statistical confidence. Our results are also relevant to the interpretation of evidence that genetic variation is being maintained by balancing selection--such evidence typically involves short loci with closely spaced linked SNPs \citep{leffler2013, charlesworth2006, segurel2012}. 
 
A topic worth further investigation is the possibility of local variation in the rate of MNMs. If most MNMs are caused by error-prone polymerase activity, it is likely that high error-prone polymerase traffic should elevate rates of MNMs and simple point mutations in the same genomic regions. Both MNMs and point mutations in these regions might be subject to elevated transversion rates, and it will be important to separate the two classes of mutations to accurately study local variation of the transition/transversion ratio as in \cite{seplyarskiy2012}. Seplyarskiy, \emph{et al}. reported that the transition/transversion ratio $\kappa$ appears depressed in the neighborhood of all human SNPs, even transitions, but we found that the apparent depression of $\kappa$ in the neighborhood of transitions disappears when SNPs in perfect LD are excluded from the analysis (Supplementary Figure S7).

MNMs have the potential to accelerate evolution by quickly changing several amino acids within a single gene \citep{schrider2011}. Our results indicate that they also have the potential to increase both sequence homogeneity and A/T content. There is evidence that repetitive sequences experience more indels and point mutations than sequences of higher complexity \citep{mcdonald2011}, possibly due to the recruitment of error-prone polymerases, giving MNM extra potential to speed up local sequence evolution by triggering downstream mutations. We are hopeful that more details about this process can be elucidated by studying the spatial and allelic distribution of MNMs. In this way, population sequencing data could provide new information about the biochemistry of replication \emph{in vivo}, providing a way to measure the activity of Pol $\zeta$ over evolutionary time. Pol $\zeta$ is tightly regulated in embryonic and adult cells because over- and under-expression can each be harmful; excess error-prone DNA replication increases the genomic mutation rate, but impaired translesion synthesis ability can lead to replication fork stalling, DNA breakage, and translocations that are more harmful than point mutations \citep{waters2006, waters2009, ogawara2010, northam2010, lange2011}. An important avenue for future work will be to assess whether different eukaryotes incur different levels of MNM because of changing evolutionary pressures being exerted on error-prone DNA replication activity throughout the tree of life.

% You may title this section "Methods" or "Models". 
% "Models" is not a valid title for PLoS ONE authors. However, PLoS ONE
% authors may use "Analysis" 
\section*{Methods}
\textbf{Data summary and accession:}

We performed all of our analyses on SNP calls that were generated by the 1000 Genomes Project Consortium using joint genotype calling on 2--6x whole genome coverage of 1,092 humans sampled from Europe and Africa \citep{1000genomes2012}. All sequences were mapped to the human reference hg19. To determine ancestral alleles, we downloaded alignments of hg19 to the primate genomes PanTro2 (chimpanzee), ponAbe2 (orangutan), and rheMac3 (rhesus macaque) from the UCSC Genome Browser. 

\textbf{Ascertainment of SNP pairs from the 1000 Genomes Phase I data:}

Let $S(L)$ be a count of SNPs that are polymorphic in a pair of haplotypes and lie $L$ bp apart with their derived alleles on the same haplotype. Similarly, let $D(L)$ be a count of SNPs with derived alleles that lie on opposite haplotypes. To measure $S(L)$ and $D(L)$ precisely from the 1000 Genomes data, we used a stringent procedure for ancestral identification, utilizing only sites that had the same allele present in chimp, orangutan, and rhesus macaque. For each pair $p$ of SNPs $L$ bp apart satisfying this criterion and passing the 4-gamete test (to avoid confounding effects of recombination and sequencing error), we counted the number of haplotypes $N_{\textrm{AA}}(p)$ carrying the ancestral allele at both sites, the number $N_{\textrm{AD}}(p)$ carrying the ancestral allele at only the first site, the number $N_{\textrm{DA}}(p)$ carrying the ancestral allele at only the second site, and the number $N_{\textrm{DD}}(p)$ with both derived alleles. Singletons are excluded because they cannot be phased. Combining this information across the set $P(L)$ of SNP pairs $L$ bp apart, we obtain counts 
\begin{equation}
S(L)=\sum_{p\in P(L)} N_{\textrm{AA}}(p)\times N_{\textrm{DD}}(p)
\end{equation}
and 
\begin{equation}
D(L)=\sum_{p\in P(L)} N_{\textrm{AD}}(p)\times N_{\textrm{DA}}(p)
\end{equation}
as desired.

The quantity $S(L)-D(L)$ has been used as an estimate of the number of MNMs lying $L$ bp apart. Since two simultaneous mutations should always lie in perfect LD, $S(L)-D(L)$ should in theory always be smaller than the following count of perfect-LD same lineage pairs:
\begin{equation}
S_{\textrm{LD}}(L)=\sum_{p\in P(L)} N_{\textrm{AA}}(p)\times N_{\textrm{DD}}(p)\times\mathbf{1}(N_{\textrm{AD}}=N_{\textrm{DA}}=0)
\end{equation}
To count perfect LD mutation pairs in a way that is more robust to genotype and phasing error, we instead compute $S_{\textrm{LD}}(L)$ as follows:
\begin{eqnarray}
S_{\textrm{LD}}(L)&=&\sum_{p\in P(L)} N_{\textrm{AA}}(p)\times N_{\textrm{DD}}(p) \\ \label{relaxed_LD}
 && \hspace{1 cm} \times\mathbf{1}\left(\frac{N_{\textrm{AD}}+N_{\textrm{DA}}}{1092}<\min\left(0.02, \frac{2N_{\textrm{AA}}+N_{\textrm{AD}}+N_{\textrm{DA}}}{2\times 2184},   \frac{N_{\textrm{AD}}+N_{\textrm{DA}}+2N_{\textrm{DD}}}{2\times 2184}\right)\right). \notag
\end{eqnarray}
This criterion is designed such that genotyping/phasing error up to 2\% will not disrupt perfect LD, but such that very low or high frequency alleles will not be considered in perfect LD unless at least half of the minor alleles appear in the same lineages.

We use a slightly different procedure to obtain the counts $N^{\textrm{LD}}_{\textrm{ts}}(L)$, $N^{\textrm{LD}}_{\textrm{m}}(L)$, and $N^{\textrm{LD}}_{\textrm{tv}}(L)$ that do not need to be compared to $S(L)-D(L)$. After dividing $P(L)$ into transition pairs, mixed pairs, and transversion pairs to obtain sets $P_{\textrm{ts}}(L), P_{\textrm{m}}(L)$, and $P_{\textrm{tv}}(L)$, we simply count the number of pairs with derived alleles occur in the exact same set of lineages:
\begin{equation}
N^{\textrm{LD}}_t(L)=\sum_{p\in P_t(L)} \textbf{1}(N_{\textrm{AD}}(p)=N_{\textrm{DA}}(p)=0)
\end{equation}
for each $t\in\{\textrm{ts},\textrm{m},\textrm{tv}\}.$ Nearby singletons are considered to be in perfect LD if the derived alleles occur in the same diploid individual.  It is this counting procedure that we use to obtain the site frequency spectra of perfect LD SNPs shown in Figure~\ref{SFS_vs_LD}. 

\textbf{Simulating SNP pairs in LD under the coalescent:}

The simulated data used to generated Figure~\eqref{LD_mixture_model} was produced using Hudson's \texttt{ms} \citep{hudson2002}. We simulated 2,184 human haplotypes (1,092 African and 1,092 European) under the demographic model published in  \citep{harris2013} that was previously inferred from tracts of identity by state in the 1000 Genomes trios. Because we were only interested in SNP pairs separated by 100 bp or less, we simulated a total of $5.6\times 10^5$ independent ``chromsomes" of length 10 kb using the mutation rate $2.5\times 10^{-8} \ \textrm{bp}^{-1}\textrm{gen}^{-1}$ and the recombination rate $1.0\times 10^{-8} \ \textrm{bp}^{-1}\textrm{gen}^{-1}$.

\textbf{Estimating the contribution of MNM to new point mutations:}

In the 1000 Genomes data, we counted $N_{\textrm{SNP}}=17,140,039$ non-singleton SNPs that met our criterion for ancestral identifiability. For each pair type $t$, we also counted the number $N_t^{\textrm{relaxed-LD}}(L)$ of $t$-type SNP pairs $L$ bp apart that met the relaxed definition of perfect LD given in equation~\eqref{relaxed_LD}. We estimate the fraction $f_{\textrm{MNM}}=0.019$ produced by MNM using the following equation: 

\begin{equation}
f_{\textrm{MNM}}=\frac{2}{N_{\textrm{SNP}}}\sum_{t\in\{\textrm{ts},\textrm{m},\textrm{tv}\}}\sum_{L=1}^{100}N_t^{\textrm{relaxed-LD}}(L)\times m_t(L)
\end{equation}

This fraction is a lower bound because it discounts singletons and MNMs spanning more than 100 bp.

\textbf{Calculating $r^2$ with a correction for multinucleotide mutation:} 

Given two SNPs $s_A,s_B$ with major alleles $A,B$ and minor alleles $a,b$, let $p_{AB}, p_{Ab}, p_{aB},$ and $p_{ab}$ be population frequencies of each of the four associated haplotypes. Let $p_A,p_a,p_B,$ and $p_b$ be the allele frequencies at individual loci. One measure of linkage disequilibrium between the loci is the correlation coefficient
$$r^2(s_A,s_B)=\frac{|p_{AB}p_{ab}-p_{aB}p_{Ab}|}{\sqrt{p_Ap_ap_Bp_b}}.$$
LD decays as a function of the genetic distance between loci. It is often useful to summarize the rate of this decay by computing the average value of $r^2(s,s')$ over all SNP pairs $(s,s')$ that occur $L$ bp apart. Letting $S(L)$ denote this set of SNP pairs, we define
$$r^2(L)=\frac{1}{|S(L)|}\sum_{(s_1,s_2)\in S(L)} r^2(s_1,s_2).$$
To avoid averaging together the effects of MNM and linked independent mutation, it would be ideal to replace $S(L)$ with the of SNPs LD bp apart that were produced by independent pairs of mutations. 

Although it is not possible to classify a SNP pair in perfect LD as an MNM unambiguously, we can correct for MNM by estimating the probability that each observed SNP $s$ was generated as part of a pair of simultaneous mutations. This probability, $\mathbb{P}_{\textrm{MNM}}(s)$, is calculated as a function of the nearest SNP $s_{\textrm{LD}}$ occurring in perfect LD with $s$. If $s$ is not in perfect LD with any other SNP within 1000 bp, we assume that $s$ was generated by an ordinary point mutation and let $\mathbb{P}_{\textrm{MNM}}(s)=0$. Otherwise, letting $A(s,s_{\textrm{LD}})$ denote the allelic state of the pair $(s,s_{\textrm{LD}})$ (either transitions (ts), transversions (tv), or mixed (m)) and $L$ denote the distance between $s$ and $s_{\textrm{LD}}$, we estimate that $\mathbb{P}_{\textrm{MNM}}(s)=m_{A(s,s_{\textrm{LD}})}(L).$
Note that when $s_1$ and $s_2$ are in perfect LD and mutually closer to one another than to any other SNP in perfect LD, 
$$\frac{1}{2}(\mathbb{P}_\textrm{MNM}(s_1)+\mathbb{P}_\textrm{MNM}(s_2))=\mathbb{P}_\textrm{MNM}(s_1)=\mathbb{P}_\textrm{MNM}(s_2).$$

After estimating $\mathbb{P}_{\textrm{MNM}}(s)$ for each SNP $s$ that occurs in $S(L)$, we use these values to compute a weighted average $r^2_{\textrm{MNM}}(L)$ that downweights each SNP by the probability that it is part of a complex mutation pair: 

$$r^2_{\textrm{MNM}}(L)=\frac{\sum_{(s_1,s_2)\in S(L)} r^2(s_1,s_2)(1-(\mathbb{P}_{\textrm{MNM}}(s_1)+\mathbb{P}_{\textrm{MNM}}(s_2))/2)}{\sum_{(s_1,s_2)\in S(L)} 1-(\mathbb{P}_{\textrm{MNM}}(s_1)+\mathbb{P}_{\textrm{MNM}}(s_2))/2}.$$

% Do NOT remove this, even if you are not including acknowledgments
\section*{Acknowledgments}
We thank Jasper Rine for sharing his deep understanding of replication and mutation in eukaryotes. We also thank members of the Nielsen and Slakin laboratories, particularly Josh Schraiber, Mehmet Somel, and Tyler Linderoth, for helpful discussion and comments. An earlier manuscript was much improved thanks to input from Michael Lynch, Molly Przeworski, and two anonymous reviewers. K.H. was supported by a National Science Foundation Graduate Research Fellowship and R.N. received funding from NIH grant 2R14003229-07.

\section*{Disclosure Declaration}

The authors have no conflicts of interest to declare.

%\section*{References}
% The bibtex filename
\bibliography{mybib}

\section*{Figure Legends}

\section*{Tables}

\end{document}

% Title must be 150 characters or less
\begin{flushleft}
{\Large
\textbf{Supplemental Information: \\
Error-prone polymerase activity causes multinucleotide mutations in humans }
}
% Insert Author names, affiliations and corresponding author email.

Kelley Harris
and
Rasmus Nielsen

\end{flushleft}

\renewcommand{\figurename}{Supplementary Figure S\hspace{-0.125 cm}}
\setcounter{figure}{0}

\renewcommand{\tablename}{Supplementary Table S\hspace{-0.125 cm}}

\begin{figure}[!ht]
\begin{center}
\includegraphics[width=0.95\textwidth]{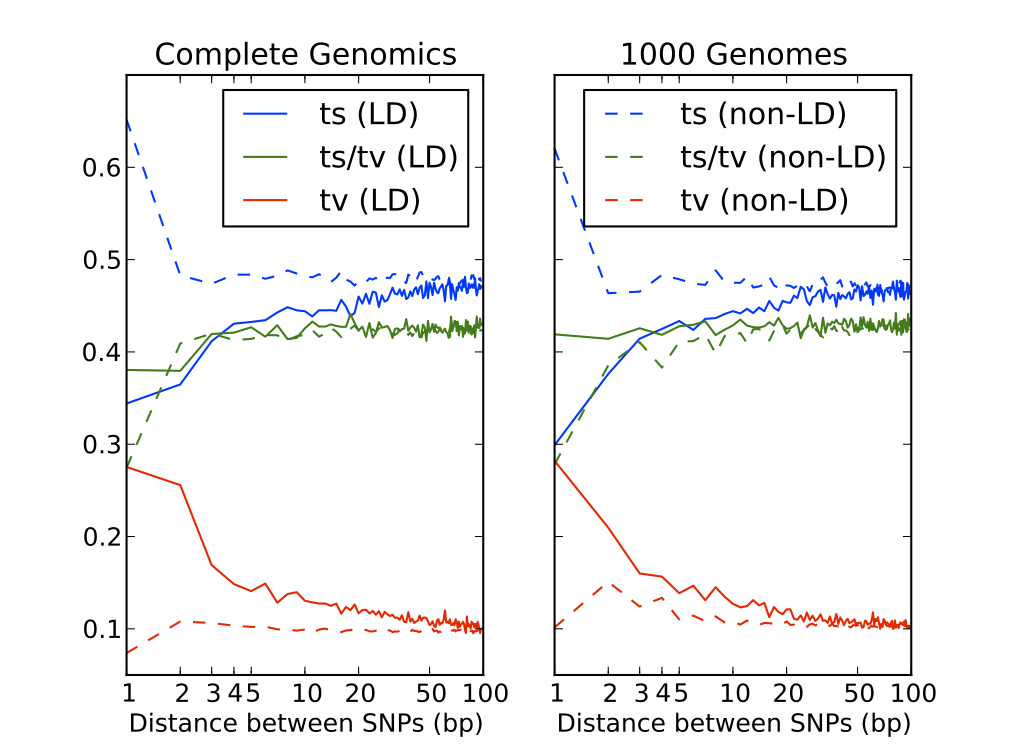}
\end{center}
\caption{
{\bf Consistency of the transition: transversion ratio across linked SNPs from different sequencing platforms.} This figure shows that excess transversions in perfect LD are not an artifact of Illumina sequencing or the 1000 Genomes pipeline, but are also present in a set of 54 human genomes sequenced by Complete Genomics (CG). To make this comparison, we subsampled 54 genomes from the 1000 Genomes Phase I dataset that had approximately the same population breakdown as the 54 CG individuals. Because the CG data are unphased, we ignore all 1000 Genomes phasing information, classifying each SNP pair as being in perfect LD if it is in perfect LD with respect to at least one possible haplotype phasing. We ignore all CG SNPs at which more than 10\% of the samples have a missing genotype. Note that most pairs of nearby SNPs in the CG data are not annotated as ``SNPs" in the MasterVarBeta files that are publicly availably, but as ``complex" substitutions where a string of two or more bases is regarded as one polymorphic unit. We ignored all complex substitutions that included indels, but extracted SNPs from each substitution multi nucleotide substitution where all variant alleles had the same length and a one-to-one mapping between sites was possible.}
\label{comp_genom_ratios}
\end{figure}

\newpage

\begin{figure}[!ht]
\begin{center}
\includegraphics[width=0.95\textwidth]{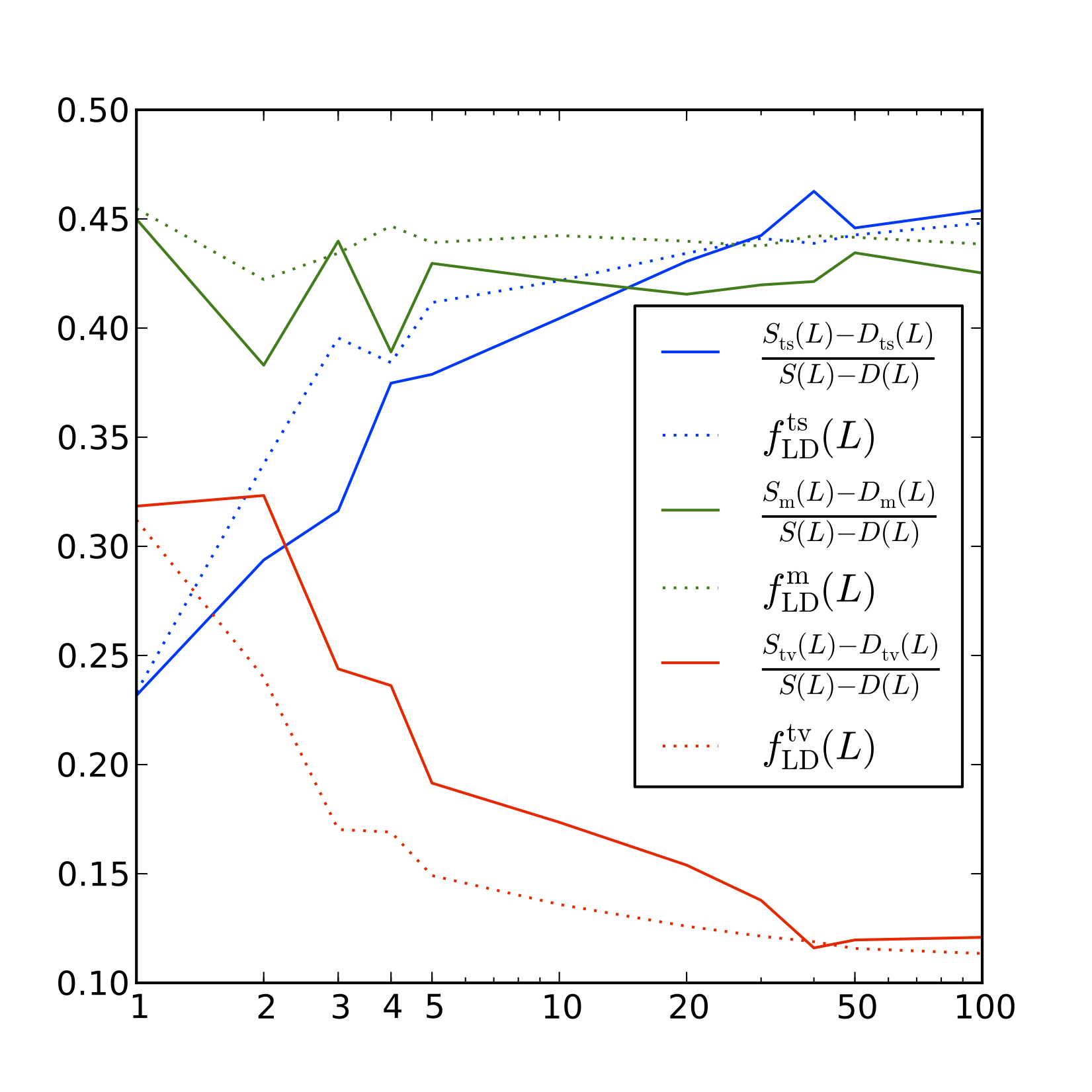}
\end{center}
\caption{
{\bf Quantifying simultaneous transitions vs. transversions.} This figure plots the relative abundances of transitions, transversions, and mixed pairs as fractions of the quantity $S(L)-D(L)$. The transversion fraction $(S_{\textrm{tv}}(L)-D_{\textrm{tv}}(L))/(S(L)-D(L))$ is slightly higher than $f_{\textrm{LD}}^{\textrm{tv}}(L)$, especially for small $L$ where MNMs are the most apparent.}
\label{compare_transv_fraction}
\end{figure}

\newpage

\begin{table}[!ht]
\begin{center}
\caption{\bf{Quantifying MNMs spanning $>100$ bp}}\label{farapart_MNMs}
\begin{tabular}{|c|cc|cc|cc|}
\hline
$L$ &  $S^{(\textrm{LD})}_{\textrm{ts}}(L)$ & $m_{\textrm{ts}}(L)$ & $S^{(\textrm{LD})}_{\textrm{m}}(L)$ & $m_{\textrm{m}}(L)$   & $S^{(\textrm{LD})}_{\textrm{tv}}(L)$  & $m_{\textrm{tv}}(L)$ \\
\hline
100 & 4.507487e+10 & 0.370434497853 & 4.414249e+10 & 0.354296958936 & 1.142233e+10 & 0.389186681775\\
200 & 4.217310e+10 & 0.379874606701 & 4.100119e+10 & 0.372376712439 & 1.054684e+10 & 0.37707655264\\
300 & 3.961781e+10 & 0.395451459315 & 3.880264e+10 & 0.377681881936 & 9.884130e+09 & 0.371605021077\\
400 & 3.807976e+10 & 0.399251605723 & 3.726721e+10 & 0.395012183525 & 9.351356e+09 & 0.38983897119\\
1000 & 2.094000e+10 & 0.426248637331 & 2.096024e+10 & 0.423787575103 & 5.275246e+09 & 0.397717995479\\
3000 & 1.287797e+10 & 0.443566715331 & 1.270760e+10 & 0.418373624368 & 3.234083e+09 & 0.382000444317\\
5000 & 9.147849e+09 & 0.403365299761 & 9.088510e+09 & 0.392177623168 & 2.377678e+09 & 0.38602646942\\
10000 & 5.147443e+09 & 0.290639194365 & 5.139715e+09 & 0.280151830098 & 1.330389e+09 & 0.267209538362\\
\hline
\end{tabular}
\end{center}
\begin{flushleft}
For each distance $L$ listed above, we counted all SNPs in perfect LD between $L$ and $L+100$ bp apart. For each pair type $t$, we subsampled haplotype pairs in order to calculate $S_t(L), D_t(L), S^{(\textrm{LD})}_t(L),$ and $m_t(L)$ aggregated over this 100 bp window. The results suggest that the ratio of MNMs to perfect LD SNPs achieves its minimum value around $L=100$ and then stops decreasing with the distance between SNPs.
\end{flushleft}
\end{table}

\newpage

\begin{figure}[!ht]
\begin{center}
\includegraphics[width=0.95\textwidth]{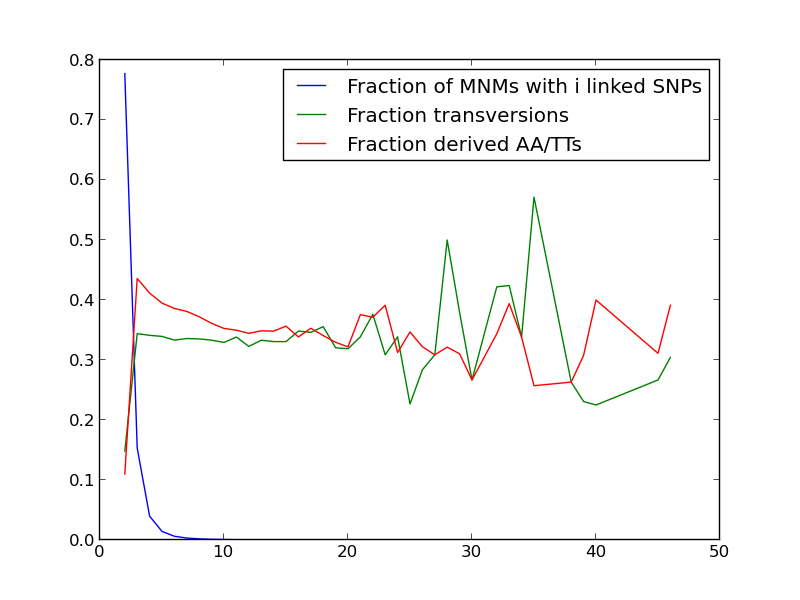}
\end{center}
\caption{
{\bf Clusters of 2 or more perfect LD SNPs.} In the 1000 Genomes data, we found all clusters of 2 or more perfect LD SNPs with fewer than 1 kb between adjacent pairs. We plot the resulting distribution of cluster sizes and compare it to the distribution of cluster sizes in data simulated under the \cite{harris2013} model using \texttt{ms}. The cluster sizes from real data are slightly more dispersed toward very small and very large clusters. It is possible that the longest clusters formed by error-prone replication of single-stranded DNA following double-strand breakage as proposed by \cite{roberts2012}.}
\label{higher_order_MNMs}
\end{figure}

\newpage

\begin{figure}[!ht]
\begin{center}
\includegraphics[width=0.95\textwidth]{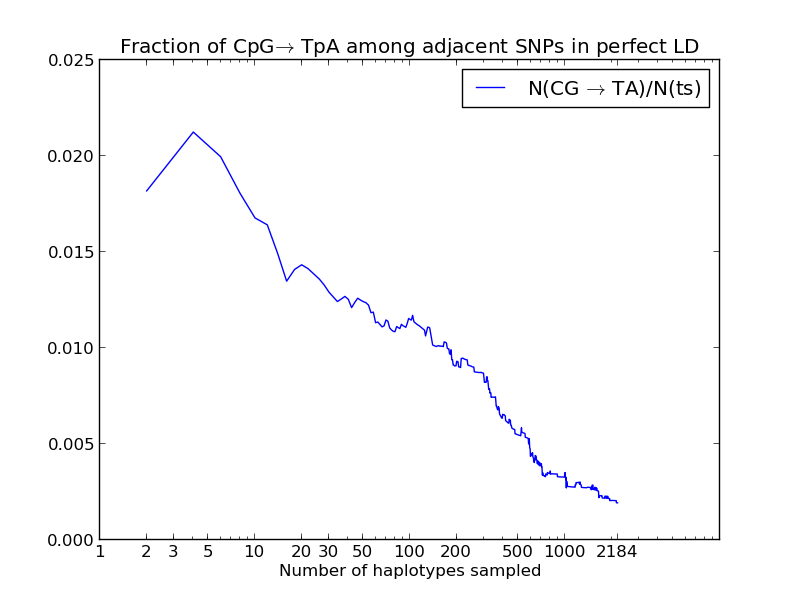}
\end{center}
\caption{
{\bf CpG double mutations in perfect LD.} It is well known that CpG dinucleotides undergo a high rate of C$\to$T mutation related to deamination of the methylated cytosine. This process should elevate the rate at which CpG mutations undergo double transitions into the TpA dinucleotide. When the CG $\to$ TA double mutation is mediated by cytosine deamination, the two mutations should occur non-simultaneously in general, and are more likely to be found in perfect LD in small samples than large samples. We counted dinucleotide mutations in subsets of the 1000 Genomes data ranging in size from 2 to 2,184 haplotypes and tabulated the fraction of perfect LD transition pairs that were of allelic type CG $\to$ TA. As shown here, this fraction declines almost 10-fold as the sample size increases from 2 to 2,184.}
\label{CpG_decline}
\end{figure}

\newpage

\begin{figure}[!ht]
\begin{center}
\includegraphics[width=0.95\textwidth]{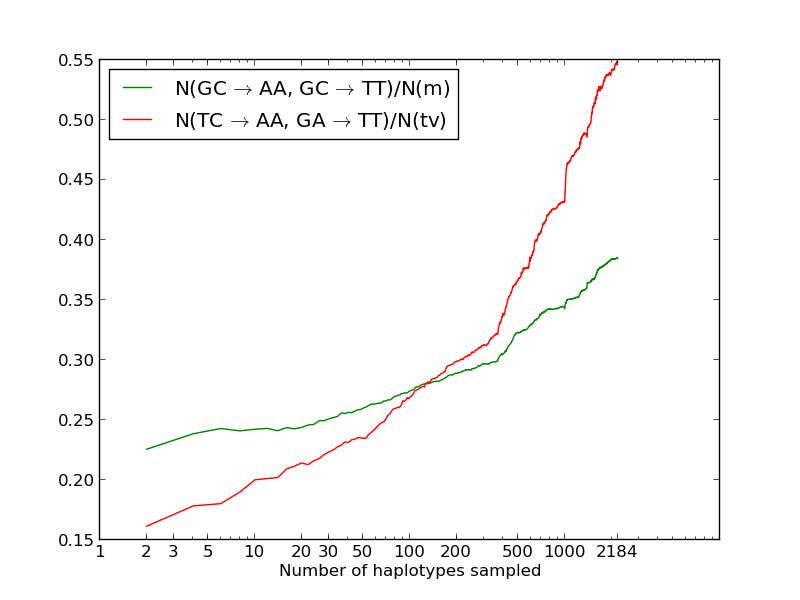}
\end{center}
\caption{
{\bf Pol $\zeta$ mutations in subsets of the 1000 Genomes data.} We counted all four types of Pol $\zeta$-associated mutations in subsamples of 2--2184 1000 Genomes haplotypes and pooled counts of the reverse complements TC$\to$AA, GA$\to$TT and GC$\to$AA, GC$\to$TT.  We divided these counts, respectively, by the total counts of adjacent transversions and adjacent mixed pairs in perfect LD. The fraction of each Pol $\zeta$-associated mutation increases with sample size as expected for MNMs.}
\label{polzeta_subsample}
\end{figure}

\newpage

\begin{figure}[!ht]
\begin{center}
\includegraphics[width=0.95\textwidth]{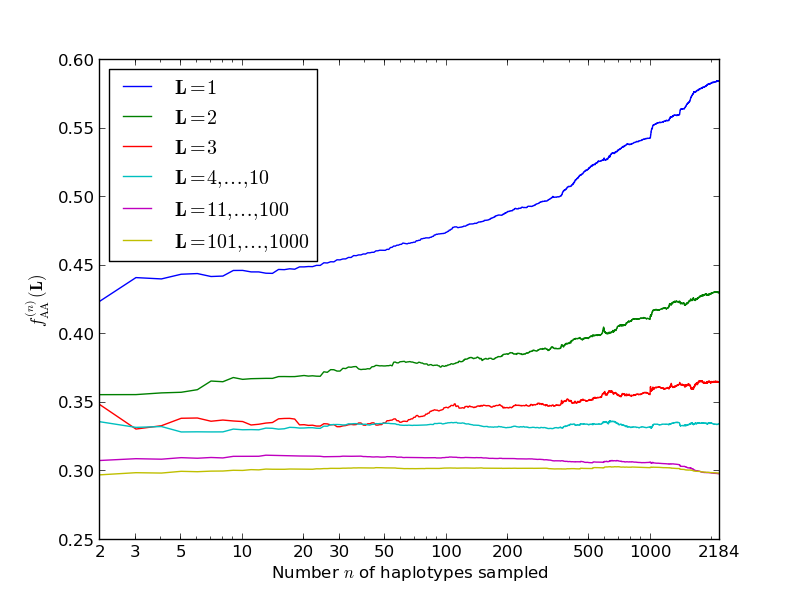}
\end{center}
\caption{
{\bf Linked derived AA/TT allele pairs.} To calculate $f_{\textrm{AA}}^{(n)}(\mathbf{L})$ for a vector $\mathbf{L}$ of lengths $L_1,\ldots,L_k$ measured in base pairs, we sampled $n$ haplotypes from the 1000 Genomes data and counted to total number of perfect LD SNPs separated by a distanced falling in the range $L_1,\ldots,L_k$. We then counted the number of these with AA or TT as the pair of derived alleles and computed the ratio of AA/TT pairs to total pairs. As shown, the three curves $f_{\textrm{AA}}^{(n)}(1),$ $f_{\textrm{AA}}^{(n)}(2),$ and $f_{\textrm{AA}}^{(n)}(3)$ are all increasing functions of $n$, the proportion of AA/TT derived alleles increasing as more haplotypes are sampled. However the same pattern does not hold for vectors $\mathbf{L}$ of longer distances. This suggests that derived AA/TT's are only overrepresented among MNMs separated by very short distances of 1 to 3 base pairs.}
\label{AA_TT_subsample}
\end{figure}

\newpage

\begin{figure}[!ht]
\begin{center}
\includegraphics[width=0.95\textwidth]{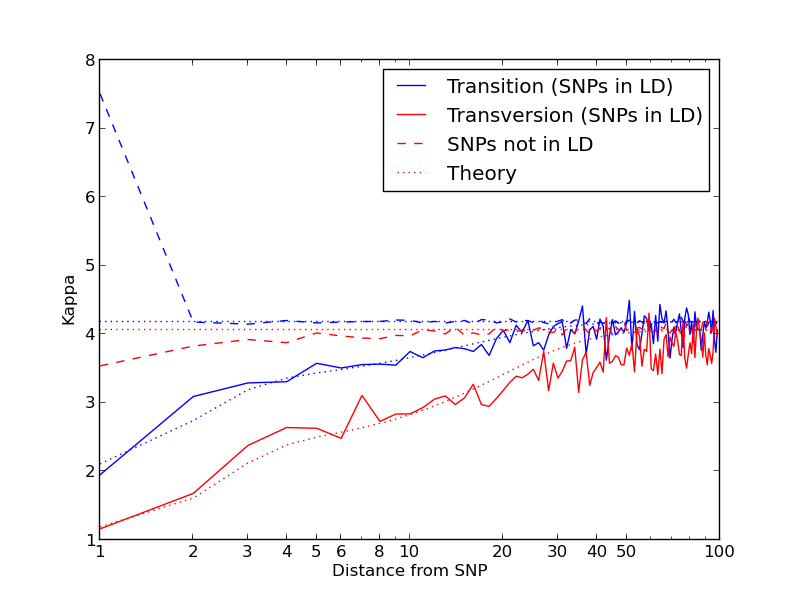}
\end{center}
\caption{
{\bf Explaining the variation of $\kappa$ in the neighborhood of a SNP.}  It is common to summarize the transition: transversion ratio with a parameter $\kappa$ for which $f_{\textrm{transversion}}=2/(\kappa+2)$. On average, $\kappa=4$ in the human genome, meaning that each type of transition is 4 times as abundant as a particular type of transversion. Looking only at SNPs in LD, $\kappa$ appears significantly depressed in the neighborhood of a segregating SNP, suggesting that a negative positional correlation between mutation rate and $\kappa$. However, this effect is not apparent among SNPs that are not in LD. $\kappa$ is elevated, on average, at the site next to a transition because of adjacent transitions generated by mutations at both positions of a CpG site. In addition, $\kappa$ appears slightly reduced among unlinked SNPs in the neighborhood of a transversion, suggesting that there is some regional variation in $\kappa$ but not as much as appears to be the case if simultaneous mutations are regarded as independent.}
\label{kappa_variation}
\end{figure}

\newpage

%\begin{table}[!ht]
%\begin{center}
%\caption{
%\bf{Exponential mixture fits to $N_{\textrm{LD}}(L)$}
%}\label{tab:three_exponentials}
%\begin{tabular}{|c|c|cc|cc|cc|cc|}
%\hline
%Number & Error & $N_1$ & $a_1$ & $N_2$ & $a_2$ & $N_3$ & $a_3$ & $N_4$ & $a_4$ \\
%of  terms & & & & & & & & & \\
%\hline
%1 & $4.916\times 10^8$ & $6.12\times 10^3$ & 0.0274 & -- & -- & -- & -- & -- & --  \\
%2 & $7.496\times 10^6$ &  $2.86\times 10^3$ & 0.00784 & $8.64\times 10^4$ & 1.33 & -- & -- & -- & -- \\
%3 & $3.759\times 10^5$ & $2.05\times 10^3$ & 0.00267 & $3.31\times 10^3$ & 0.126 & $1.18\times 10^5$ & 1.74 &  -- & -- \\
%4 & $3.759\times 10^5$ &  $2.05\times 10^3$ & 0.00267 & $3.31\times 10^3$ & 0.126 & $5.91\times 10^4$ & 1.74 & $5.91\times 10^4$ & 1.74  \\
%\hline
%\end{tabular}
%\end{center}
%\begin{flushleft}
%For $N$ ranging from 1 to 4, we fit empirical distribution $N_{\textrm{LD}}(L)$ to $\sum_{i=1}^N N_i\exp(-a_i\cdot L)$ using the least-squares curve fit function from the Python library \texttt{scipy.optimize}. This table records the minimum squared error achieved for each $N$ along with the vector of optimal parameter values. A significantly better fit is achieved by using three exponential terms rather than one or two, but no additional improvement is achieved by using four terms. These parameter values are slightly different from the values in Table~\ref{M-tab:param_estimates} of the main text because they were not fit jointly with transition/transversion abundance parameters.
%\end{flushleft}
%\end{table}

%\newpage

 \section{Exponential decay of LD over short genomic distances}\label{proof_exponential_decay}
  In data simulated under the standard coalescent with recombination using \texttt{ms} \citep{hudson2002}, we saw that the count $N_{\textrm{LD}}(L)$ of SNPs in perfect LD $L$ bp apart decays approximately exponentially for $L$ between 1 and 100 bp. Here, we give a heuristic argument why this should be true in the asymptotic limit $L\ll 1/\rho$, where $\rho$ is the population-scaled recombination rate.
 
 Let $T_1,\ldots,T_L$ be the sequence of $n$-leaf coalescence trees that occur at the sites of a sequence of length $L$ that has been evolving with mutation and recombination parameters $\theta$ and $\rho$. For simplicity, we assume a constant effective population size $N$. The rates $\theta$ and $\rho$ are population-scaled such that $\mu=\theta/(4N)$ is the mutation rate per site per generation and $r=\rho/(4N)$ is the recombination rate per site per generation. Given any of these trees $T_i$, let $P(T_i)$ be the set of points $(x,y)\in T_i$ with the property that $x$ and $y$ lie on the same branch of $T_i$. The sequential coalescent yields a natural map from points on $T_i$ to points on $T_{i+1}$, though not every point on $T_i$ necessarily maps to a point on $T_{i+1}$ if a recombination has occurred between the sites. Let $\epsilon_{x,y}(T_i)$ be defined such that $1-\epsilon_{x,y}(T_i)$ is the probability that $x$ and $y$ both map to $T_{i+1}$,  $(x,y)\in P(T_{i+1})$, and the branch containing $(x,y)$ subtends the same set of lineages in both $T_i$ and $T_{i+1}$.
 
A pair of points $(x,y)\in P(T_i)$ can give rise to a pair of SNPs in perfect LD at sites $i$ and $j$ if the following events occur: E1) a mutation occurs at position $x$ on tree $T_i$, E2) $x$ and $y$ map to a single branch of each tree between $T_i$ and $T_j$ that subtends the same set of lineages, and E3) A mutation occurs at position $y$ on tree $T_j$. Not every pair of SNPs in perfect LD must correspond to a pair of points $x,y$ satisfying E1--E3; for example, the integrity of the clade by the branch containing $x$ and $y$ could be broken up and re-formed by two separate recombinations occurring between sites $i$ and $j$. If the sample size $n$ is relatively large, however, it will be combinatorially unlikely for any clade to re-form after it has been broken up by recombination, particularly within a very short genomic window. Motivated by this, we will estimate $N_{\textrm{LD}}(L)$ assuming that all linked SNP pairs arise at pairs of points $(x,y)$ that satisfy E1--E3 for some $T_i$ and $T_j$. 

 Integrating over $x,y$ and $T_i,\ldots,T_{i+L}$, we compute that the probability of observing a pair of SNPs in perfect LD at sites $i$ and $i+L$ is the following:

\begin{eqnarray}
N_{\textrm{LD}}(i,i+L)&=&\theta^2\int_{T_i,\ldots,T_{i+L}}\int_{(x,y)\in P(T_i)}(1-\epsilon_{x,y}(T_i))\cdots (1-\epsilon_{x,y}(T_{i+L}))d_{(x,y)}d_{(T_i,\ldots,T_{i+L})}\\ \label{exact_formula}
&=&\theta^2+\theta^2\sum_{k=1}^L(-1)^k \int_{T_i,\ldots,T_{i+L}}\int_{(x,y)\in P(T_i)} \sum_{i\leq j_1<\cdots<j_k\leq i+L} \epsilon_{x,y}(T_{j_1})\cdots \epsilon_{x,y}(T_{j_k})d_{(x,y)}d_{(T_i,\ldots,T_{i+L})}.\notag
\end{eqnarray}
Let $\ell(T)$ denote the total branch length of tree $T$. Since any alteration of tree structure requires a recombination event, $\epsilon_{x,y}(T)\leq \rho\cdot \ell(T)$. This implies that
\begin{equation}
\sum_{i\leq j_1<\cdots<j_k\leq i+L} \epsilon_{x,y}(T_{j_1})\cdots \epsilon_{x,y}(T_{j_k}) \leq (\epsilon_{x,y}(T_i)+\cdots+\epsilon_{x,y}(T_{i+L}))^k\leq (L\rho(\ell(T_i)+\cdots+\ell(T_{i+L})))^k
\end{equation}
for every $k$.  Letting $T^{(2)}$ denote the sum of squares of the branch lengths of a coalescent tree $T$, this implies that  
\begin{eqnarray} 
N_{\textrm{LD}}(i,i+L) &=& \theta^2-\theta^2\int_{T_i,\ldots,T_{i+L}}\int_{(x,y)\in P(T_i)} \rho(\epsilon_{x,y}(T_i)+\cdots+\epsilon_{x,y}(T_{i+L}))d_{(x,y)}d_{T_i,\ldots,T_{i+L}}+O((\rho L)^2)\notag \\
&=&\theta^2\mathbb{E}(T^{(2)})(1-\rho L\cdot\mathbb{E}(\epsilon_{x,y}(T_i)))+O((\rho L)^2).
%&=& \theta^2 \mathbb{E}(T^{(2)})\exp( -\rho L\cdot\mathbb{E}(\epsilon_{x,y}(T_i))+O((\rho L)^2)
\end{eqnarray}
In human-like data where $N=10,000$ and $\rho=0.0004$, we can see that $\rho L\leq 0.04 \ll 1$ when $L<100$. Therefore, the first-order linear decay rate of $N_{\textrm{LD}}(i,i+L)$ is small compared to $L$. In addition, we can see from equation~\eqref{exact_formula} that the $O((\rho L)^2)$ term of the Taylor expansion will be positive, meaning that $N_{\textrm{LD}}(i,i+L)$ has concave upward shape. This makes it reasonable, for our purposes, to approximate $N_{\textrm{LD}}(i,i+L)$ by an exponential function.

%, putting us in the appropriate asymptotic realm for perfect LD to decay exponentially with distance.
 
\section{Enrichment of MNMs in large datasets}\label{asymptotics_T2}
As a consequence of the argument in Section S\ref{proof_exponential_decay}, we saw that the abundance of linked independent mutations in a sample of $n$ lineages is proportional to the expected sum of squared branch lengths in an $n$-leaf coalescence tree. This is a simple consequence of the fact that two mutations must affect a single branch to create SNPs in perfect LD. In contrast, the abundance of MNMs should be proportional to the total tree length, just as the total number of segregating sites is proportional to the expected tree length. 

It is a standard result in population genetics that the expected total tree length $\mathbb{E}(T_{\textrm{total}})$ equals the harmonic number $\sum_{i=1}^{n-1} 1/i$ \citep{watterson1975}. To show this, let $T_i$ be the length of time that the a random genealogy has exactly $i$ lineages, which has distribution function $f_i(t)=\binom{i}{2}\exp(-t\binom{i}{2}).$ It follows that 
\begin{equation}
\mathbb{E}(T_{\textrm{total}})=\mathbb{E}\left(\sum_{i=2}^n iT_i\right)=\sum_{i=2}^n i\mathbb{E}(T_i)=\sum_{i=2}^n i\cdot\frac{2}{i(i-1)}=\sum_{i=1}^{n-1}\frac{1}{i}\approx\log(n-1)
\end{equation}
Therefore, if $\mu_{\textrm{MNM}}$ is the rate of MNMs per coalescent time unit, the expected number of MNMs approaches infinity with increasing $n$ at the asymptotic rate $\mu_{\textrm{MNM}}\log(n).$

In contrast, if $\mu$ is the rate of ordinary point mutations, linked independent mutations appear at the rate $\mu^2\mathbb{E}(T_{\textrm{total}}^{(2)})$, where $T_{\textrm{total}}^{(2)}$ is the sum of squares of the coalescent tree branch lengths. We can show that $\mathbb{E}(T_{\textrm{total}}^{(2)})$ approaches a constant as $n\to\infty$. To proceed, we let $\ell_1,\ldots,\ell_n$ denote the lengths of the $n$ leaves of the tree and $b_{n-1},\ldots,b_2$ denote the lengths of the $n-2$ internal branches, indexed such that the more recent endpoint of branch $i$ is the first time when the tree has $i$ lineages:
\begin{equation}
\mathbb{E}(T_{\textrm{total}}^{(2)})=n\mathbb{E}(\ell_n^2)+\sum_{i=2}^{n-1}\mathbb{E}(b_i^2).
\end{equation}
Given that a branch is present when the tree has $i$ lineages, the probability that the branch is ended by the next coalescence event is $(i-1)/\binom{i}{2}=2/i$. Therefore, given $j<i$, the probability that $b_i=T_i+\cdots+T_j$ is 
\begin{equation}
\mathbb{P}(b_i=T_i+\cdots+T_j)= \left(1-\frac{2}{i}\right)\cdots\left(1-\frac{2}{j+1}\right)\cdot \frac{2}{j}=\frac{(i-2)\cdots (j-1)\cdot 2}{i\cdots (j+1)\cdot j}=\frac{2(j-1)}{i (i-1)}.
\end{equation}
It follows that
\begin{eqnarray}
\mathbb{E}(b_i^2)&=&\sum_{j=2}^i \mathbb{P}(b_i=T_i+\cdots+T_j)\cdot \mathbb{E}((T_i+\cdots+T_j)^2)\\
&=&\sum_{j=2}^i \frac{2(j-1)}{i (i-1)}\left(\sum_{k=j}^i \mathbb{E}(T_k^2)+2\sum_{j\leq k<\ell\leq i} \mathbb{E}(T_k)\mathbb{E}(T_\ell) \right)\\
&=&\sum_{j=2}^i \frac{2(j-1)}{i (i-1)}\left(\sum_{k=j}^i\frac{8}{k^2(k-1)^2}+\sum_{j\leq k<\ell\leq i} \frac{8}{k(k-1)\ell(\ell-1)} \right)\\
&=&\sum_{j=2}^i \frac{2(j-1)}{i (i-1)}\left(\sum_{k=j}^i\frac{4}{k^2(k-1)^2}+\left(\sum_{k=j}^i\frac{2}{k(k-1)} \right)^2\right).\\
&=&\sum_{j=2}^i \frac{2(j-1)}{i (i-1)} \left(\frac{4}{3}\left(\frac{1}{j^3}-\frac{1}{i^3} \right)+\left(\frac{2}{j-1}-\frac{2}{i} \right)^2+O\left(\frac{1}{j^4}+\frac{1}{i^4}\right)\right).\\
&=&\frac{2}{i(i-1)}\left(4\log(i-1)-3/2 \right)+O(i^{-3}).
\end{eqnarray}
This implies that
\begin{eqnarray}
\mathbb{E}(T_{\textrm{total}}^{(2)})&=&n\mathbb{E}(b_n^2)+\sum_{i=2}^{n-1}\mathbb{E}(b_i^2)\\
&=& \frac{8\log(n-1)}{n-1}+\sum_{i=2}^{n-1} \frac{8\log(i-1)}{i(i-1)}+O(1/n)\\
&=&\frac{1}{2}(\log(2)+1)+\frac{7\log(n-1)}{n-1}+O(1/n),
\end{eqnarray}
which decreases asymptotically to the limit $(\log(2)+1)/2$ as $n$ approaches infinity. 

It may seem counterintuitive that $\mathbb{E}(T_{\textrm{total}}^{(2)})$ decreases as more lineages are sampled and $\mathbb{E}(T_{\textrm{total}})$ increases unboundedly, but in both simulated and real data we observe fewer SNPs in perfect LD in a sample of 2,184 haplotypes than in a subset of e.g. 1,000 haplotypes. To explain why, we note that the total tree length grows at rate $\log(n)$ as more lineages are sampled, but the tree length is subdivided among distinct branches at the faster rate $O(1/n)$. Because branch subdivision occurs faster than the growth rate of the total tree length, the sum of squared branch lengths decreases with increasing sample size, reducing the prevalence of independent linked SNPs and enhancing the signature of MNMs.

\section{Close LD SNPs in a single diploid genome}\label{schrider_replication}

In their paper titled ``Pervasive multinucleotide mutational events in eukaryotes," Schrider, \emph{et al}. used a chimpanzee outgroup to polarize SNPs found in human trio data \citep{schrider2011}. For each pair of SNPs fewer than 20 bp apart, they recorded whether the derived alleles lay on the same phased haplotype or on different haplotypes. Figure 3 of their main text records the number of nearby derived alleles that they found on the same lineage ($N_S(L)$) or on different lineages ($N_D(L)$) as a function of the distance $L$ between the sites. By the reasoning in their paper, $N_D(L)$ should equal the number of derived allele pairs found on the same haplotype $L$ bp apart that were produced by independent mutations. $N_S(L)-N_D(L)$ should equal the number that were created by multinucleotide mutation. From visual inspection of their Figure 3, it appears that
$$\frac{N_S(1)-N_D(1)}{N_S(1)}\approx \frac{1400-2000}{1400}\approx 0.86,$$
meaning that about 86\% of adjacent same-lineage SNPs were caused by MNM. In contrast, in appears that 
$$\frac{N_S(20)-N_D(20)}{N_S(20)}<\frac{100}{2000}=0.05,$$
meaning that fewer than 5\% of same-lineage SNPs 20 bp apart were caused by MNM.

Schrider, \emph{et al.} defined a multinucleotide polymorphism (MNP) to be a pair of SNPs within 20 bp of each other where both derived alleles lie on the same haplotype of a diploid genome that has been phased using trio information. For each possible set of two ancestral and two derived alleles, they tabulated the abundance of that pair as a fraction of all MNPs in their dataset. In particular, they reported that GC$\to$TT plus its reverse complement GC$\to$AA made up 2.3\% of the total, more than any other mixed transition/transversion mutation pair but fewer than the 16\% of adjacent perfect LD mutations reported in our paper. Similarly, TC$\to$AA and GA$\to$TT made up 1.5\% of their total, compared to 10\% of ours. Derived AA/TT allele pairs made up 15.8\% of their total MNPs. 

We believe that two factors explain the differences in these results. One is that Schrider, \emph{et al}, pooled together all MNPs separated by 1--20 bp instead of focusing on adjacent polymorphisms. The other is that they were only able to sample pairs in perfect LD in a sample of size two, which should contain more linked independent mutations than perfect LD pairs in a larger sample of haplotypes. To verify this, we replicated the Schrider, \emph{et al}. analysis on our data, tabulating all pairs of SNPs less than 20 bp apart where the derived alleles lay on the same haplotype of at least one 1000 Genomes individual. We obtained that GC$\to$TT plus GC$\to$AA made up 3.0\% of the total, TC$\to$AA and GA$\to$TT made up 12.8\%, and XY$\to$AA plus XY$\to$TT made up 19.0\%. These numbers are similar enough that the two datasets do not appear to be qualitatively different; it is only the difference in sampling scheme that reveals more Pol $\zeta$-associated mutations in our analysis of the 1000 Genomes data. Figure S\ref{polzeta_subsample} confirms that Pol $\zeta$-associated tandem mutations increase in frequency as more lineages are sampled, suggesting that these mutations are enriched among MNMs. Figure S\ref{AA_TT_subsample} demonstrates a similar result for the frequency $f_{\textrm{AA}}(L)$ of derived AA/TT allele pairs in perfect LD at distances of $L=$1--3 bp apart. Although $f_{\textrm{AA}}(L)$ decreases as $L$ increases from 4 to 1,000, it does not appear to depend on the number of haplotypes sampled. A possible explanation is that MNMs only create excess derived AA/TTs that lie 1--3 bp apart but that transcriptional strand bias creates excess AA/TTs among independent mutations that occur at nearby sites \citep{green2003}.

\section{Disruption of MNMs by recombination}\label{breakup}
We have seen that MNM is not the only source of perfect LD SNPs in genetic data. Conversely, not all MNMs need give rise to perfect LD SNPs; it is possible for recombination to decouple two derived alleles that have been created by a single complex mutation. To assess how many MNMs are likely to be broken down in this way, we used \texttt{ms} to simulate many short ancestral recombination graphs on 2,184 haplotypes. To mimic the composition of the 1000 Genomes data, we sampled 492 African haplotypes and 1,692 non-African haplotypes according to the model of human history proposed in \cite{harris2013}. For each length $L$ listed in the table below ($L=1$, 10, 100, 1000, 10000), we simulated LOOKUP sequences $L$ bp long assuming a constant recombination rate of $1.0\times 10^{-8}$ recombinations per site per generation. Given the trees at the left and right ends of each simulated sequence $s_i$, we calculated the probability $p_i$ that a point placed on the left tree uniformly at random would be ancestral to a subtree that had not recombined between the left tree and the right tree. Letting $\ell_i$ be the total branch length of the leftmost tree, we obtain the following estimate $\hat{\rho}_{\textrm{MNM}}(L)$ for the probability that an MNM spanning $L$ bp will be broken up by recombination: 
\begin{equation}
\hat{\rho}_{\textrm{MNM}}(L)= \frac{\sum_{i=1}^k (1-p_i)\cdot \ell_i}{\sum_{i=1}^k \ell_i}
\end{equation}
The resulting values of $\hat{\rho}_{\textrm{MNM}}(L)$ are recorded in Table~\ref{tab:recombination}. Even for $L=10000$, we estimate that only 22\% of MNMs will recombine out of perfect LD. Although 10,000 bp is about ten times the mean recombination distance for a sample of two lineages, the majority of MNMs are expected to occur at low frequency on the sample of 2,184 haplotypes and persist in LD for relatively long genomic distances. 

\begin{table}[!ht]
\begin{center}
\caption{
\bf{Estimates of $\hat{\rho}_{\textrm{MNM}}(L)$}
}\label{tab:recombination}
\begin{tabular}{|c|c|}
\hline
$L$ & $\hat{\rho}_{\textrm{MNM}}(L)$ \\
\hline
1 & $6.60\times 10^{-5}$ \\
10 & $3.20\times 10^{-4}$ \\
100 & $3.64\times 10^{-3}$ \\
1000 & $4.70\times 10^{-2}$ \\
10000 & 0.224 \\
\hline
\end{tabular}
\end{center}
\begin{flushleft}
\end{flushleft}
\end{table}

\section{Simulating data with a realistic MNM distribution}\label{how_to_simulate}

We argue that MNM affects many features of genetic data including SNP density, the local transition/transverion ratio, and linkage disequilibrium. To capture these effects, it may be useful for readers to incorporate MNM into simulations of human-like SNP data. Tables~S\ref{tab:simulation_params} and~S\ref{tab:simulation_params2} provide a framework for doing this. For each SNP pair type $t\in\{\mathrm{ts},\mathrm{m},\mathrm{tv}\}$ and each distance $L$ between 1 and 100 bp, the table entry $(L,t)$ provides the probability $P_{\textrm{MNM}}(L,t)$ that a given mutation should be an MNM of type $t$ and spacing $L$. The entries of the two tables add up to $0.018$, indicating that MNMs should account for 1.8\% SNPs. To simulate a dataset with $\theta$ total SNPs, one should first use a program such as \texttt{ms} to generate a dataset with $\theta\times (1-0.009)$ total SNPs. After this, one should select a fraction $P(L,t)$ of SNPs uniformly at random to be MNMs of type $t$ and spacing $L$. For each selected SNP, a new SNP should be introduced in perfect LD exactly $L$ bp to the left.

\begin{table}[!ht]
\begin{center}
\caption{
\bf{Table of Parameters for Simulating Data with Realistic MNM (Part 1 of 2)}}\label{tab:simulation_params}
\begin{tabular}{|c|ccc|}
\hline
$L$ & ts & m & tv \\
\hline
1 & 0.000689075099285 & 0.00139395307314 & 0.0011649961804 \\
2 & 0.000197253881722 & 0.000264153610395 & 0.000248926345734 \\
3 & 9.24985667775e-05 & 0.00014275347624 & 9.31164214363e-05 \\
4 & 0.000114899830562 & 0.000122700855385 & 8.44464230223e-05 \\
5 & 9.3567275539e-05 & 0.000114868148911 & 5.64888107995e-05 \\
6 & 8.36026776582e-05 & 9.12884447692e-05 & 4.88456557619e-05 \\
7 & 7.58303670667e-05 & 8.55603852686e-05 & 4.15000412291e-05 \\
8 & 7.5266572516e-05 & 8.47592580657e-05 & 4.42345400698e-05 \\
9 & 7.59109091453e-05 & 8.5119765307e-05 & 4.01596006209e-05 \\
10 & 7.11589265039e-05 & 7.91113112853e-05 & 3.86046895154e-05 \\
11 & 7.1486593824e-05 & 7.26941774735e-05 & 3.21955518884e-05 \\
12 & 7.18145139791e-05 & 7.38269722864e-05 & 3.14879573414e-05 \\
13 & 6.48462106821e-05 & 7.07801448587e-05 & 3.13363299385e-05 \\
14 & 6.77155120397e-05 & 7.02332783973e-05 & 3.01738531827e-05 \\
15 & 6.52151208566e-05 & 6.85536171231e-05 & 2.86070366858e-05 \\
16 & 6.83303623306e-05 & 6.734269853e-05 & 2.8152154477e-05 \\
17 & 6.53380909148e-05 & 7.12488875399e-05 & 2.9415716168e-05 \\
18 & 6.28786897511e-05 & 7.09363924191e-05 & 2.97695134415e-05 \\
19 & 6.72646218263e-05 & 6.64833369478e-05 & 2.62820831742e-05 \\
20 & 6.77155120397e-05 & 6.56239753657e-05 & 2.74951023976e-05 \\
21 & 6.31325919102e-05 & 6.26737000383e-05 & 2.31857258277e-05 \\
22 & 6.11743849651e-05 & 6.08755305214e-05 & 2.15897676445e-05 \\
23 & 6.07827435761e-05 & 6.29359330929e-05 & 2.1678431988e-05 \\
24 & 6.13310415207e-05 & 6.08380686565e-05 & 2.07917885529e-05 \\
25 & 5.98428042425e-05 & 6.06132974668e-05 & 2.20774215338e-05 \\
26 & 5.9607819409e-05 & 6.27860856331e-05 & 2.20774215338e-05 \\
27 & 5.99211325203e-05 & 6.26737000383e-05 & 2.07917885529e-05 \\
28 & 5.9646983548e-05 & 6.1100301711e-05 & 2.14567711293e-05 \\
29 & 6.15660263542e-05 & 5.95269033837e-05 & 2.03041346636e-05 \\
30 & 5.91378497422e-05 & 5.92272084642e-05 & 1.97278164308e-05 \\
31 & 6.44333909867e-05 & 5.74757746569e-05 & 1.85649923763e-05 \\
32 & 6.63962625002e-05 & 6.4106112176e-05 & 1.74589928305e-05 \\
33 & 6.53721556235e-05 & 5.87093258232e-05 & 1.7182492944e-05 \\
34 & 6.69509870583e-05 & 5.95573922501e-05 & 1.77354927169e-05 \\
35 & 6.23851772334e-05 & 6.09836857862e-05 & 1.76169927656e-05 \\
36 & 6.53294845037e-05 & 5.74372261829e-05 & 1.57209935442e-05 \\
37 & 6.21718216341e-05 & 5.77070655006e-05 & 1.74984928143e-05 \\
38 & 6.10623725178e-05 & 5.92490044585e-05 & 1.83674924574e-05 \\
39 & 6.20438082746e-05 & 5.53170601158e-05 & 1.67084931387e-05 \\
40 & 6.1019701398e-05 & 5.7360129235e-05 & 1.71429929603e-05 \\
41 & 6.30261621992e-05 & 6.22107504952e-05 & 1.78347642869e-05 \\
42 & 6.02395698522e-05 & 5.98087910166e-05 & 1.65308074641e-05 \\
43 & 6.15509074273e-05 & 6.00489869645e-05 & 1.82974586434e-05 \\
44 & 5.97478182616e-05 & 6.0849640124e-05 & 1.89704686165e-05 \\
45 & 6.00756526554e-05 & 5.89280725412e-05 & 1.68673124506e-05 \\
46 & 5.79857083952e-05 & 6.11699013878e-05 & 1.74561961771e-05 \\
47 & 6.17967832226e-05 & 6.05694115182e-05 & 1.76244486703e-05 \\
48 & 6.07313214429e-05 & 6.04493135443e-05 & 1.88863423699e-05 \\
49 & 5.90921494741e-05 & 6.06094441762e-05 & 1.65728705874e-05 \\
50 & 6.16328660257e-05 & 5.76870601439e-05 & 1.72038174371e-05 \\
\hline
\end{tabular}
\end{center}
\end{table}

\begin{table}[!ht]
\caption{
\bf{Table of Parameters for Simulating Data with Realistic MNM (Part 1 of 2)}}\label{tab:simulation_params2}
\begin{center}
\begin{tabular}{|c|ccc|}
\hline

51 & 6.57879112884e-05 & 6.09082009503e-05 & 1.91108170597e-05 \\
52 & 6.11257758428e-05 & 6.17346351559e-05 & 1.98371188957e-05 \\
53 & 6.45360415855e-05 & 6.04536621373e-05 & 1.82029397647e-05 \\
54 & 6.03487532685e-05 & 6.13214180531e-05 & 1.89746354654e-05 \\
55 & 5.97875702982e-05 & 6.30569298848e-05 & 1.74766379287e-05 \\
56 & 6.41906982191e-05 & 6.0412340427e-05 & 1.87476661417e-05 \\
57 & 6.44497057439e-05 & 6.11974529223e-05 & 1.74766379287e-05 \\
58 & 6.31115001993e-05 & 6.05776272681e-05 & 1.74766379287e-05 \\
59 & 6.1082607922e-05 & 6.0412340427e-05 & 1.67957299574e-05 \\
60 & 6.30683322785e-05 & 5.73132121561e-05 & 1.94285741129e-05 \\
61 & 5.92263873279e-05 & 5.90074022775e-05 & 1.87476661417e-05 \\
62 & 6.06940966349e-05 & 5.80983246514e-05 & 1.72042747402e-05 \\
63 & 6.05214249517e-05 & 6.07842358195e-05 & 1.81121520352e-05 \\
64 & 6.24208134666e-05 & 5.88421154364e-05 & 1.81121520352e-05 \\
65 & 6.2334477625e-05 & 5.95445845111e-05 & 1.74312440639e-05 \\
66 & 6.07804324765e-05 & 6.086687924e-05 & 1.78397888467e-05 \\
67 & 6.32841718825e-05 & 6.09082009503e-05 & 1.62510035804e-05 \\
68 & 5.97012344566e-05 & 6.20238871278e-05 & 1.74766379287e-05 \\
69 & 5.9010547724e-05 & 6.01644101653e-05 & 1.68865176869e-05 \\
70 & 6.01329136646e-05 & 5.83875766233e-05 & 1.90654231949e-05 \\
71 & 6.22049738626e-05 & 5.91313674084e-05 & 1.85206968179e-05 \\
72 & 5.97012344566e-05 & 6.06189489784e-05 & 1.63871851747e-05 \\
73 & 5.95285627735e-05 & 5.84288983336e-05 & 1.72042747402e-05 \\
74 & 6.14711192092e-05 & 5.99991233242e-05 & 1.74766379287e-05 \\
75 & 6.11689437636e-05 & 5.68999950533e-05 & 1.87022722769e-05 \\
76 & 5.99170740606e-05 & 6.19412437073e-05 & 1.72042747402e-05 \\
77 & 6.11257758428e-05 & 5.77264292589e-05 & 1.69773054164e-05 \\
78 & 5.88810439616e-05 & 5.87594720158e-05 & 1.89746354654e-05 \\
79 & 6.20754701003e-05 & 5.66520647916e-05 & 1.72950624697e-05 \\
80 & 6.04350891101e-05 & 5.80570029411e-05 & 1.65233667689e-05 \\
81 & 6.16006229715e-05 & 5.73958555767e-05 & 1.83845152237e-05 \\
82 & 5.90537156447e-05 & 5.97511930625e-05 & 1.65687606337e-05 \\
83 & 6.28093247537e-05 & 5.92140108289e-05 & 1.77943949819e-05 \\
84 & 6.16869588131e-05 & 5.63628128197e-05 & 1.70226992812e-05 \\
85 & 5.6290968714e-05 & 5.83875766233e-05 & 1.72950624697e-05 \\
86 & 5.72406629714e-05 & 6.02057318756e-05 & 1.77490011172e-05 \\
87 & 6.01760815854e-05 & 6.0123088455e-05 & 1.81575458999e-05 \\
88 & 6.12552796052e-05 & 5.98751581934e-05 & 1.61148219862e-05 \\
89 & 6.06940966349e-05 & 5.87181503056e-05 & 1.72496686049e-05 \\
90 & 5.97444023774e-05 & 5.76024641281e-05 & 1.72042747402e-05 \\
91 & 6.11689437636e-05 & 5.75198207075e-05 & 1.77943949819e-05 \\
92 & 5.78881817833e-05 & 5.69413167636e-05 & 1.64325790394e-05 \\
93 & 6.03487532685e-05 & 5.63628128197e-05 & 1.67503360927e-05 \\
94 & 5.63341366348e-05 & 5.78503943897e-05 & 1.57062772034e-05 \\
95 & 6.1298447526e-05 & 5.67347082122e-05 & 1.58878526624e-05 \\
96 & 6.03055853477e-05 & 5.95859062214e-05 & 1.65233667689e-05 \\
97 & 5.84925326744e-05 & 5.70239601842e-05 & 1.81575458999e-05 \\
98 & 5.96148986151e-05 & 5.7147925315e-05 & 1.57062772034e-05 \\
99 & 5.93990590111e-05 & 5.71066036047e-05 & 1.66595483632e-05 \\
100 & 6.23776455458e-05 & 5.73132121561e-05 & 1.55247017444e-05 \\
\hline
\end{tabular}
\end{center}
\begin{flushleft}
\end{flushleft}
\end{table}

\bibliography{mybib}